%% file: ms.tex
\documentclass[aps, prb, twocolumn, superscriptaddress,floatfix]{revtex4-2}
\include{preamble}
\include{preamble_for_diff}

\begin{document}

\title{
  Stable-to-unstable transition in quantum friction
}

\author{Daigo Oue}
\email{daigo.oue@gmail.com}
\affiliation{Instituto de Telecomunica\c{c}\~{o}es, Instituto Superior T\'{e}cnico, University of Lisbon, 1049-001 Lisbon, Portugal}
\affiliation{The Blackett Laboratory, Imperial College London, London SW7 2AZ, United Kingdom}
\author{J. B. Pendry}
\affiliation{The Blackett Laboratory, Imperial College London, London SW7 2AZ, United Kingdom}
\author{M\'{a}rio G. Silveirinha}
\affiliation{Instituto de Telecomunica\c{c}\~{o}es, Instituto Superior T\'{e}cnico, University of Lisbon, 1049-001 Lisbon, Portugal}

\date{\today}

\input{src/abstract.tex}

\maketitle

\input{src/introduction.tex}
\input{src/doppler-dielectric.tex}
\input{src/stability.tex}
\input{src/noqf.tex}
\input{src/qf.tex}
\input{src/discussion.tex}
\input{src/conclusion.tex}

\input{src/acknowledgement.tex}

\appendix
\input{src/appx_EE.tex}
\input{src/appx_reciprocity.tex}
\input{src/appx_green.tex}
\input{src/appx_F.tex}
\input{src/appx_consistency.tex}

\bibliography{%
  all,%
  textbook_all%
}
\end{document}

%% file: preamble.tex
\usepackage[]{physics}
\usepackage[]{amsmath}
\usepackage{amssymb}
\usepackage{mathrsfs}

\usepackage[T1]{fontenc}
\usepackage{fontawesome}

\usepackage[]{graphicx}
\graphicspath{{src/figures/}}
\usepackage[]{mathtools}
\usepackage[]{bm}
\usepackage[]{braket}
\usepackage{esint}
\usepackage[normalem]{ulem}
\usepackage{cancel}
\usepackage[colorlinks=true]{hyperref}
\usepackage[caption=false]{subfig}

\newcommand{\figref}[1]{\figurename~\ref{#1}}

\usepackage{cases}
\usepackage[]{comment}

\newcommand{\rec}{{} ^ \star\hspace{-.2em}}
\newcommand{\recE}{{} ^ \star\hspace{-.2em}\bm{E}}
\newcommand{\recH}{{} ^ \star\hspace{-.2em}\bm{H}}
\newcommand{\recj}{{} ^ \star\hspace{-.2em}\bm{j}}
\newcommand{\recG}{{} ^ \star\hspace{-.1em}G}

%% file: src/abstract.tex
\begin{abstract}
  We investigate the frictional force arising from quantum fluctuations when two dissipative metallic plates are set in a shear motion.
  While early studies showed that the electromagnetic fields in the quantum friction setup reach nonequilibrium steady states, yielding a time-independent force, other works have demonstrated the failure to attain steady states, leading to instability and time-varying friction under sufficiently low-loss conditions.
  Here, we develop a fully quantum-mechanical theory without perturbative approximations and unveil the transition from stable to unstable regimes of the quantum friction setup.
  Due to the relative motion of the plates, their electromagnetic response may be active in some conditions, resulting in optical gain. We prove that the standard fluctuation-dissipation leads to inconsistent results when applied to our system, and, in particular, it predicts a vanishing frictional force. Using a modified fluctuation-dissipation relation tailored for gain media, we calculate the frictional force in terms of the system Green's function, thereby recovering early works on quantum friction.
  Remarkably, we also find that the frictional force diverges to infinity as the relative velocity of the plates approaches a threshold. This threshold is determined by the damping strength and the distance between the metal surfaces. Beyond this critical velocity, the system exhibits instability, akin to the behaviour of a laser cavity, where no steady state exists. In such a scenario, the frictional force escalates exponentially. Our findings pave the way for experimental exploration of the frictional force in proximity to this critical regime. 
\end{abstract}

%% file: src/introduction.tex
\section{Introduction}
Time-varying media have recently attracted the attention and curiosity of researchers across various disciplines, 
ranging from optics \cite{galiffi2022photonics} and acoustics \cite{shen2019nonreciprocal,wen2022unidirectional,xu2020physical} to condensed matter physics \cite{oka2009photovoltaic,mciver2020light, matsuo2013mechanical,kobayashi2017spin}.
One of the popular classes of time-varying media is the travelling-wave type.
In these systems, their intricate spatiotemporal variations evoke the electrodynamics of moving dielectrics.
In the domain of optics, for instance, phenomena such as Fresnel drag~\cite{huidobro2019fresnel}, Čerenkov emission~\cite{oue2022cerenkov,oue2023noncontact}, and nonreciprocal wave propagation~\cite{galiffi2019broadband} have elicited substantial scholarly exploration.
These analogies, particularly in the low-energy regime, have been underscored by the homogenisation theories~\cite{huidobro2019fresnel,huidobro2021homogenization,serra2023rotating,serra2023homogenization,prudencio2023replicating}.

Within this context, it is worth revisiting electromagnetic phenomena within systems in motion, a particular instance of a time-variant platform.
Specifically, in this article, we study noncontact quantum frictional forces that emerge between two surfaces in relative motion.

Early studies~\cite{annett1987long,brevik1993friction,hoye1992friction,schaich1981dynamic,pendry1997shearing,pendry1998can,volokitin1999theory,volokitin2007near} of quantum friction have unveiled a mechanism rooted in elementary excitations induced by the sliding motion of surfaces and the ensuing momentum transfer facilitated by these excitations.
This motion-induced friction-effect establishes interesting connections with phenomena such as \v{C}erenkov radiation~\cite{cherenkov1934visible}, the dynamical Casimir effect~\cite{moore1970quantum,wilson2011observation,lahteenmaki2013dynamical}, Zel'dovich superradiance~\cite{zel1971generation,zel1972amplification} and its analogs in acoustics~\cite{faccio2019superradiant,Cromb2020amplification}, magnonics~\cite{wang2022twisted}, and cold atomic physics~\cite{berti2022superradiant}, and even Hawking radiation~\cite{hawking1974black}.
Moreover, in the context of field-mediated momentum transfer, quantum friction is closely related to Coulomb drag~\cite{gramila1991mutual,persson1998theory,volokitin2011quantum,narozhny2016coulomb}, wherein electron momentum transfer occurs through the mediation of electromagnetic fields between closely spaced leads.
Even loss-free dielectrics can, in principle, give rise to friction: sufficiently large shear velocities will cause them to emit light~\cite{pendry1998can}.
The light is emitted in the form of correlated photon pairs: one photon into each of the two lossless dielectrics~\cite{pendry1998can}.
Radiative heat transfer is also linked to quantum friction, as the electromagnetic fields serve as mediators for the transfer of energy between two bodies at different temperatures without any direct physical contact ~\cite{polder1971theory,joulain2005surface,volokitin2007near,biehs2021near}.

Past investigations have shown that the electromagnetic field within the quantum friction setup attains nonequilibrium steady states, resulting in constant drag forces and system stability. 
Conversely, other works ~\cite{silveirinha2013quantization,silveirinha2014optical,silveirinha2014spontaneous,silveirinha2014theory,brevik2022fluctuational} have shown that, under sufficiently low-loss conditions, the system fails to attain steady states, thus creating a time-varying frictional force and ensuing instability.

In this work, we develop a fully quantum-mechanical theory without perturbative approximations and elucidate the transition between stable and unstable regimes in the quantum friction phenomenon.
The quantisation of electromagnetic fields can be achieved through diverse methodologies, encompassing canonical quantisation~\cite{huttner1992quantization}, geometric quantisation~\cite{schnitzer2019geometric}, the path-integral formalism~\cite{bechler1999quantum,artyszuk2003effective,difallah2019path}, and methods based on Green's function (GF)~\cite{gruner1995correlation,gruner1996green,dung1998three,scheel1998qed,raabe2007unified}.
In the present case, the GF-based method is suitable, as it encapsulates the information on photonic structure within the dyadic Green's function.
The GF-based quantisation has also been widely applied to calculate quantum optical phenomena in complex geometries, including the Casimir forces~\cite{sambale2008van,buhmann2004casimir,sambale2009impact,amooghorban2011casimir,soltani2017nonmonotonic}, thermal radiation~\cite{rodriguez2013fluctuating,polimeridis2015fluctuating,khandekar2019circularly}, luminescence from quantum emitters~\cite{scheel1999quantum,hummer2013weak,feist2020macroscopic,sanchez2020cumulant,medina2021few,sanchez2022theoretical}, superfluorescence~\cite{yokoshi2017synchronization}, radiative energy transfer~\cite{joulain2005surface}.
By analysing the transition to the unstable regime from the perspective of Green's functions and the corresponding equations of motion, we will establish that the instabilities in the quantum friction setup bear resemblance to the Kelvin-Helmholtz (KH) instability.
The KH instability manifests at the interface between two fluids with different flow velocities and is part of a well-known family of hydrodynamic instabilities characterised by growing perturbations at interfaces, with other examples including the Rayleigh-Taylor and Richtmyer-Meshkov instabilities~\cite{chandrasekhar1961hydrodynamic,drazin_2002,charru2011hydrodynamic,schmid2012stability}.
The KH instability has also been studied within plasma physics~\cite{hamlin2013role,alves2014electron,alves2015slow}, and a series of experiments demonstrated that the instabilities show up in shear plasma flows~\cite{harding2009observation,hurricane2009high,smalyuk2012experimental,hurricane2012validation,kuramitsu2012kelvin}.

\begin{figure}[htbp]
  \centering
  \includegraphics[width=\linewidth]{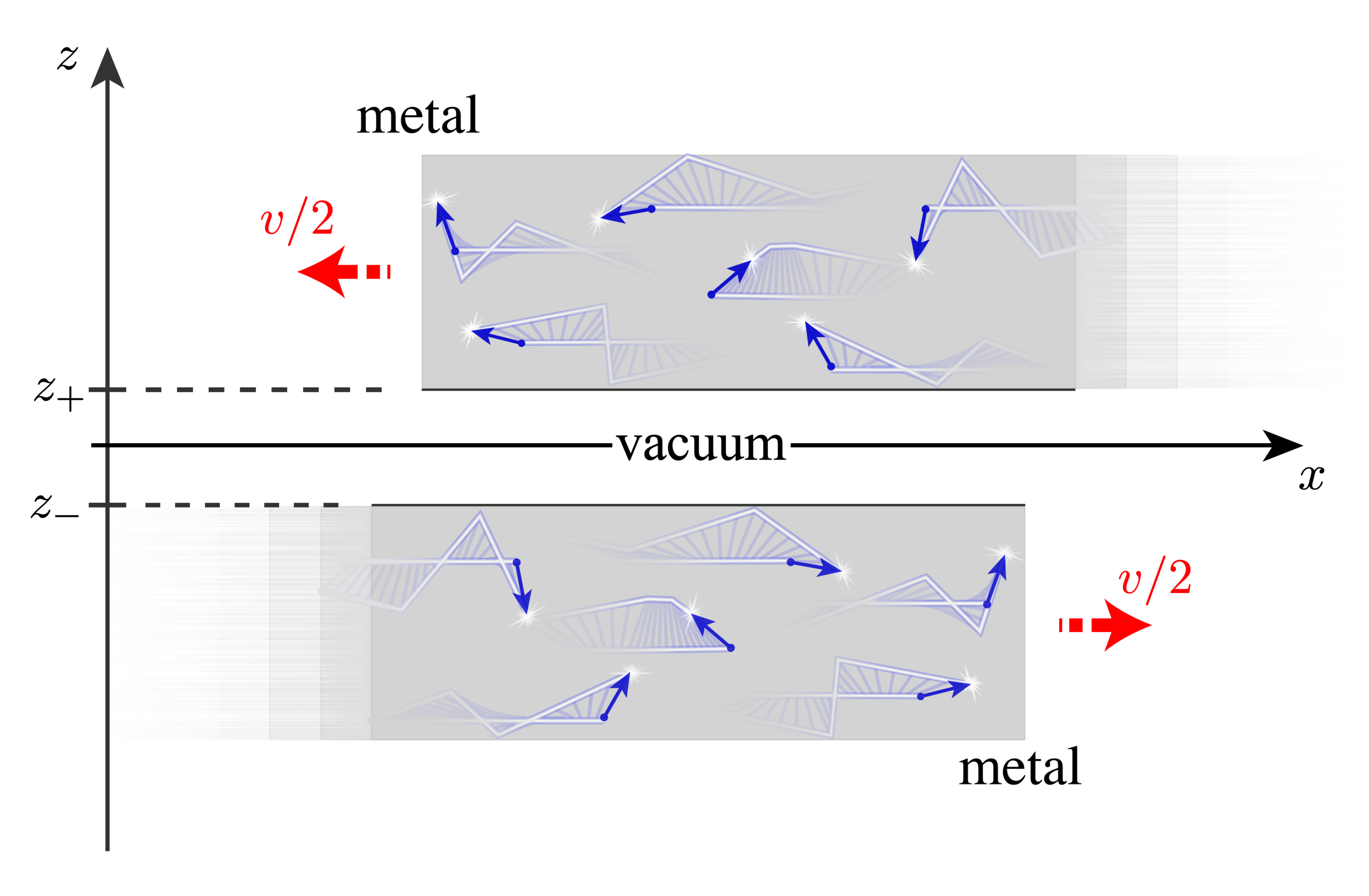}
  \caption{
    Schematic of the system analysed in this work.
    It consists of two metal sheets in a shear motion separated by a vacuum gap.
    A metallic medium occupies upper ($z > z _ +$) and lower semi-infinite ($z < z _ -$) regions.
    The distance between the surfaces is $L = z _ + - z _ -$.
    The medium filling the upper (lower) side is moving towards the left (right) at a constant speed $v/2$.
    The metallic medium is modelled by the Drude permittivity \eqref{eq:eps_pm} with a finite damping constant, which implies the existence of noise currents in the medium (indicated by blue arrows).
  }
  \label{fig:setup}
\end{figure}

Our setup consists of a metal-vacuum-metal system as shown in \figref{fig:setup}.
The metallic medium occupies the upper region ($z > z _ +$) and the lower region ($z < z _ -$), while there is a vacuum gap in between ($z _ - < z < z _ +$).
In the respective co-moving frames, the metallic slabs are described by the Drude model,
\begin{align}
  \epsilon _ \mathrm{D} (\omega) = 1 - \frac{\omega _ \mathrm{p} ^ 2}{\omega ^ 2 + i\omega \gamma},
  \label{eq:drude}
\end{align}
where $\omega _ \mathrm{p}$ and $\gamma$ are the plasma frequency and the damping constant, respectively.
In our setup, the upper and lower media are moving relative to one another at a constant velocity $v$ along the $x$ direction.
Hence, the optical response in the laboratory frame may be described by the Doppler-shifted permittivity:
\begin{align}
  \epsilon _ {\omega \bm{k} \pm} = \epsilon _ \mathrm{D}(\omega \pm k _ x v/2),
  \label{eq:eps_pm}
\end{align}
where the subscripts $+$ and $-$ specify the upper and lower slabs, respectively.
For simplicity, here we neglect the bianisotropic coupling that would arise under a Lorentz transformation~\cite{kong1986electromagnetic}. 
Note that the permittivity in the laboratory frame depends on the wave number component $k_x$ along the direction of motion.
Thus, the response is spatially dispersive.
Overall, we can write the permittivity distribution as
\begin{align}
  \epsilon _ {1} (z _ 1;v)
  = \begin{cases}
    \epsilon _ {1+} & (z _ + < z _ 1), \\
    1 & (z _ - < z _ 1 < z _ +),\\
    \epsilon _ {1-} & (z _ 1 <z _ -),
  \end{cases}
  \label{eq:eps_dist}
\end{align}
where we introduced a shorthand notation (e.g.~$\epsilon _ {1+} = \epsilon _ {\omega _ 1 \bm{k} _ 1+}$). We shall use the indices ``1'' and ``2'' to distinguish observation points from source points.
For conciseness, in the following, we omit the arguments and/or subscripts.

Since our setup has continuous translational symmetry in the $xy$ plane, we can focus on the Fourier component of the electric field,
\begin{align}
  \bm{E} _ 1(z _ 1) = \frac{1}{2\pi}\int \bm{E}(t, \bm{x} _ \parallel,z _ 1) e ^ {i\omega _ 1 t - i \bm{k} _ 1 \cdot \bm{x} _ \parallel}\dd{t}\dd{\bm{x} _ \parallel},
\end{align}
where we defined the transverse wavevector $\bm{k} _ 1 = k _ {1x} \bm{u} _ x + k _ {1y} \bm{u} _ y$ and the position vector $\bm{x} _ \parallel = x \bm{u} _ x + y \bm{u} _ y$ with the unit vector $\bm{u} _ {x(y)}$ in the $x$ ($y$) direction.
Note that we have adopted the shorthand notation, $\bm{E} _ 1(z _ 1) := \bm{E} _ {\omega _ 1, \bm{k} _ 1}(z _ 1)$.
The Fourier vector amplitude of the field satisfies a wave equation, which is derived from Maxwell's equations,
\begin{align}
  \qty[\bm{\mathcal{D}} _ 1 \times \bm{\mathcal{D}} _ 1 \times - \frac{\omega _ 1 ^ 2}{c ^ 2} \epsilon _ 1] \bm{E} _ 1 = i\omega _ 1 \mu _ 0 \bm{j} _ 1,
\end{align}
where $\bm{\mathcal{D}} _ {1} := i\bm{k} _ 1 + \bm{u} _ z \pdv*{z _ 1}$ is the Fourier-transformed gradient operator, $\mu _ 0$ is the vacuum permeability, $\bm{j} _ 1$ is the electric current density, and the arguments have been omitted for simplicity.
In our problem, the currents are due to quantum or thermal fluctuations.
Thus, $\bm{j}$ stands for a noise current operator with zero expectation value, $\expval{\bm{j} _ 1} = 0$.
The electric field generated by the noise current is determined by the system's Green's function, as follows:
\begin{align}
  \bm{E} _ 1 = i\int G _ {12} \omega _ 2 \mu _ 0\bm{j} _ 2 ^ + \dd{2} + \mathrm{H.c.},
  \label{eq:E=Gj}
\end{align}
where we have defined the positive-frequency part of the fluctuating current [i.e., $\bm{j} _ 2 = \bm{j} _ 2 ^ +\;(\omega _ 2 > 0)$], following the standard phenomenological quantisation procedure of macroscopic quantum optics~\cite{gruner1995correlation,gruner1996green,dung1998three,scheel1998qed,raabe2007unified}, introduced a shorthand notation for the integral measure, $\dd{2} = \dd{\omega _ 2}\dd{\bm{k} _ 2}\dd{z _ 2}/4\pi^2$, with the index ``2'' denoting a source point.
Note that the frequency integral should be limited to the positive frequency domain, $\omega _ 2 > 0$. The symbol ``H.c.'' stands for the Hermitian conjugate operator.
The Green's function $G$ satisfies
\begin{align}
  \qty[\bm{\mathcal{D}} _ 1 \times \bm{\mathcal{D}} _ 1 \times - \frac{\omega _ 1 ^ 2}{c ^ 2} \epsilon _ 1] G _ {12} = \delta _ {12},
\end{align}
where by definition $\delta _ {12} = \delta(\bm{k} _ 1 - \bm{k} _ 2)\delta(\omega _ 1 - \omega _ 2)\delta(z _ 1 - z _ 2)$, and we applied the shorthand notation [i.e., $G _ {12} = G _ {\omega_1,\bm{k} _ 1,\omega _ 2,\bm{k} _ 2}(z _ 1, z _ 2)$].
Note that $\delta _ {12}$ includes the term $\delta(\bm{k}_1 - \bm{k}_2)\delta(\omega_1 - \omega_2)$ in its definition. Thus, our Green's function definition differs from the standard one by a multiplication factor determined by that term.
We shall discuss in more detail the properties of the fluctuating current $\bm{j}$ in the subsequent sections.

The fluctuation-induced force can be found from the quantum expectation of Maxwell's stress tensor.
For simplicity, in this article, we shall adopt a quasi-static approximation such that the Maxwell field is dominated by the electric field.
Thus, the frictional force acting on the lower metal slab is determined by the following symmetrised field correlation function:
\begin{align}
  T _ {12} = 
  \frac{1}{2}\epsilon _ 0 \expval{\acomm{\bm{E} _ 1}{\bm{E} _ 2}}, 
\end{align}
where we introduced the vacuum permittivity $\epsilon _ 0$ and defined the anti-commutation relation $\acomm{\bm{E} _ 1}{\bm{E} _ 2} = \bm{E} _ 1 \bm{E} _ 2 + \bm{E} _ 2 \bm{E} _ 1$, and the fields are evaluated in the vacuum region immediately above the lower metal slab.
Note that the magnetic part will not contribute to the $xz$ (frictional) component in the quasi-static limit, where the field is predominantly electric because retardation effects are negligible and the magnetic response of the materials is trivial.
It was previously shown that the quasi-static approximation compares well with the exact relativistic analysis when the distance between the metal slabs is subwavelength~\cite{silveirinha2014optical}.

%% file: src/doppler-dielectric.tex
\section{Doppler-induced wave amplification in dielectrics}
\label{sec:doppler-dielectric}
Let us first discuss the dielectric response of our system before considering the fluctuating source.
We focus on the imaginary part of the response function that controls wave dissipation in the system.

If the material bodies are at rest, due to causality and passivity, the imaginary part of a dielectric function is positive for positive frequencies, $\epsilon''(\omega) > 0\ (\omega > 0)$, leading to dissipation.
As the fields are real-valued, the dielectric function satisfies $\epsilon (-\omega ^ *) = \epsilon ^ *(\omega)$; 
hence, the imaginary part of the permittivity is negative for negative frequencies, $\epsilon'' < 0\ (\omega < 0)$.
Note that the double prime represents the imaginary part of a complex number (e.g., $\epsilon = \epsilon' + i\epsilon''$).
The Drude model~\eqref{eq:drude} is consistent with the enunciated properties.

When the dielectric material moves with a constant velocity in the reference frame of interest, the situation can change substantially.
Indeed, a moving dielectric potentially emits electromagnetic radiation and cannot be regarded as a passive material~\cite{pendry1998can,silveirinha2013quantization,silveirinha2014theory,silveirinha2014spontaneous,silveirinha2014optical}.
In fact, the physical motion allows for the exchange of kinetic energy and electromagnetic energy.
In particular, the material can give away some of its kinetic energy to the electromagnetic field, behaving thus as a gain medium.
Note that the light pressure exerts some force on the material, and this implies (due to the nonzero velocity of the body) a variation of the kinetic energy of the body.
The energy exchange can be bi-directional (i.e., in some cases, it leads to additional dissipation, whereas in other cases, it leads to light emission).
The described property is well predicted by the Doppler shifted permittivity~\eqref{eq:eps_pm}.
In fact, it can be readily checked that the imaginary part of the Doppler shifted permittivities can be \textit{negative} for positive frequencies. For the upper medium, the relevant spectral range that leads to gain is determined by:
\begin{align}
  \epsilon _ {\omega \bm{k} +}'' = \epsilon _ \mathrm{D}''(\omega + k _ x v/2) < 0
  \quad (0 < \omega < \frac{\abs{k _ x} v}{2},\ k _ x < 0),
  \label{eq:eps_p<0}
\end{align}
whereas for the lower medium, it is determined by,
\begin{align}
  \epsilon _ {\omega \bm{k} -}'' = \epsilon _ \mathrm{D}''(\omega - k _ x v/2) < 0
  \quad (0 < \omega < \frac{\abs{k _ x} v}{2}, k _ x > 0).
  \label{eq:eps_m<0}
\end{align}
It can be shown that including the bianisotropic response of the moving material leads to qualitatively similar conclusions~\cite{silveirinha2014optical}.
When $\epsilon''<0$ for a positive $\omega$, the electromagnetic waves oscillating with that frequency and wave number may be amplified.
This scenario does not violate the material passivity, as the shearing of the two slabs allows for the realization of mechanical work. Consequently, this mechanical action may enable electromagnetic waves to draw energy from the system's mechanical degrees of freedom.
\begin{figure}[tbp]
  \centering
  \includegraphics[width=\linewidth]{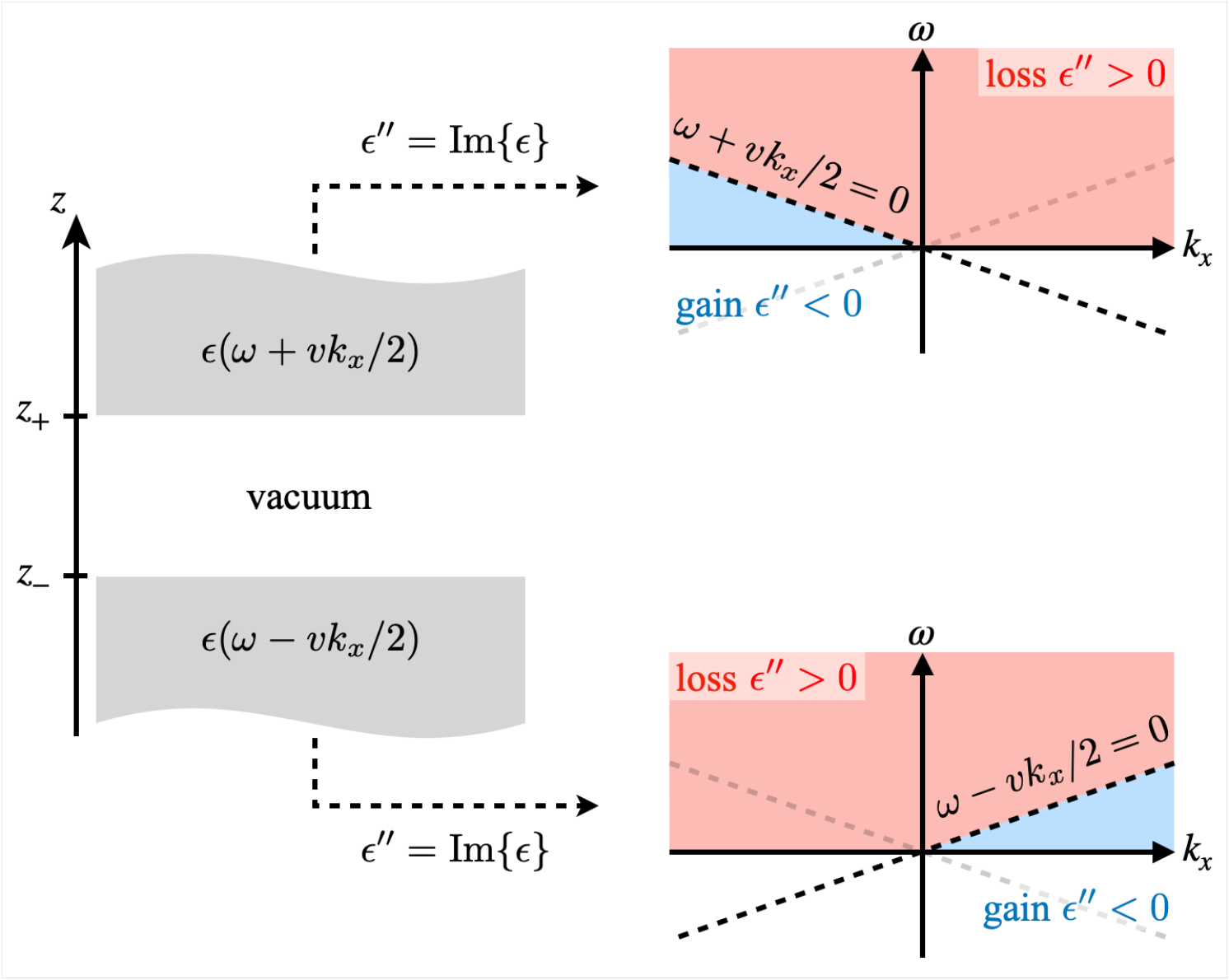}
  \caption{
    Direction-selective absorption of electromagnetic waves with short wavelengths by sheared dielectrics.
    For short wavelengths, $\omega < \abs{k _ x}v/2$, the upper (lower) medium is active (dissipative) for negative wavenumbers ($k _ x < 0$), and dissipative (active) for positive wavenumbers ($k _ x > 0$). In the former case, 
    the waves are eventually absorbed by the lower medium, whereas in the latter case, they are absorbed by the upper medium.
    Each medium receives momenta opposite to their motion and experiences a drag force.
  }
  \label{fig:direction-selective-absorption}
\end{figure}
Note that the upper region behaves as a ``gain'' medium for negative $k_x$, whereas the lower region behaves as a dissipative medium in the same spectral region (see \figref{fig:direction-selective-absorption}).
This implies that the short-wavelength left-propagating waves may be amplified in the upper region and are eventually absorbed by the lower region.
Thus, the lower medium (moving to the right) acquires momenta directed to the left direction so that it experiences a drag force.
A similar discussion holds true for waves with a sufficiently large positive $k_x$.
In this case, right-propagating waves are amplified by the lower slab and damped by the top slab.
Thus, the bottom region also experiences a force that acts to reduce its momentum.
This direction-selective absorption mechanism results in a frictional force between the two surfaces.
Note that wave amplification due to the optical gain provided by the linear motion is only feasible for short-wavelength waves, such as surface plasmons.
The discussion above is closely related to the Zeldovich superradiance~\cite{zel1971generation,zel1972amplification}, where rotational (instead of linear) motion and the rotational Doppler shift play roles.
The wave amplification by the Zeldovich mechanism has been extensively discussed and experimentally verified~\cite{faccio2019superradiant,gooding2020reinventing,Cromb2020amplification,wang2022twisted,berti2022superradiant}.
It is also beneficial to note that the optical gain by the linear motion of a metallic plate can be mimicked by electric currents driven by a dc bias~\cite{volokitin2011quantum,shapiro2017fluctuation,morgado2018drift}.

%% file: src/stability.tex
\section{Stability}
\begin{figure*}[htbp]
  \centering
  \includegraphics[width=.9\linewidth]{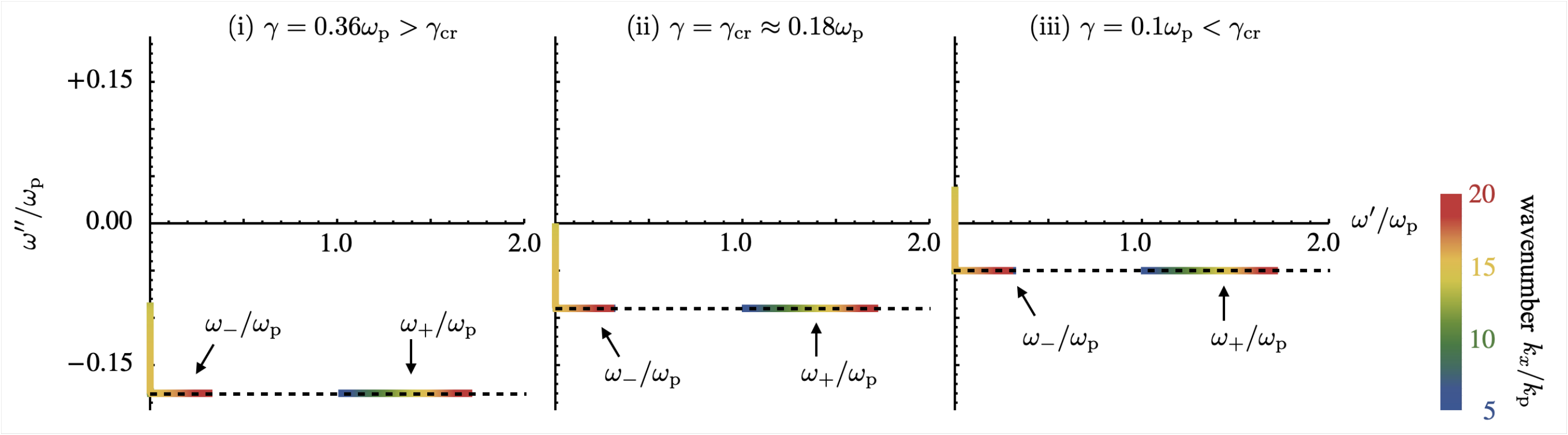}
  \caption{Roots of the characteristic equation~\eqref{eq:dispersion} for various wavenumbers $k _ x$ in the (i) stable, (ii) critical, and (iii) unstable regimes.
    The diagram shows two branches of roots represented by colored lines, positioned around the imaginary axis ($\Re\qty{\omega} = 0$) and along the horizontal line, $\Im\qty{\omega} = -\gamma/2$.
    The low-frequency root moves towards the imaginary axis as the wavenumber $k _ x$ increases.
    Once it reaches the imaginary axis, it starts to move upwards and enters the upper half side ($\Im\qty{\omega} > 0$) of the complex frequency plane.
    This indicates a mode exponentially growing in time, rather than damped, and the system will face instabilities.
    The black dashed line corresponds to $\Im\qty{\omega} = -\gamma/2$.
    The following parameters were used to generate the plots: 
    $v = 0.1c$, $L = 0.1 /k _ \mathrm{p}$, and $k _ y = 0$.
  }
  \label{fig:poles}
\end{figure*}
As discussed in the previous section, the moving slabs may behave as gain media; hence, the system may exhibit a stable-to-unstable transition if the gain overcomes the dissipation.
Here, we analyse the stability of the system.

The stability of our system is controlled by the characteristic equation~\cite{silveirinha2014spontaneous,silveirinha2014optical},
\begin{align}
  1 - r _ + r _ - = 0,
  \label{eq:dispersion}
\end{align}
where $r _ \pm$ are the reflection coefficients for the upper and lower surfaces.
Here, for simplicity, we adopt the reflection coefficients for the quasi-static regime ($\left| {{k_x}} \right| \gg \omega /c$),
\begin{align}
    r _ \pm = \frac{1 - \epsilon _ \pm}{1 + \epsilon _ \pm}e ^ {\mp2\abs{\bm{k}}z _ \pm}.
    \label{eq:r _ pm}
\end{align}
Note that the frequency and wavenumber labels are omitted for simplicity ($\epsilon _ \pm := \epsilon _ {\omega \bm{k}\pm}$).
The roots of this equation correspond to the poles of the system Green's function.
In the lossless limit, the characteristic equation~\eqref{eq:dispersion} gives the dispersion relation~\cite{brevik2022fluctuational},
\begin{align}
  \omega _ \pm = \sqrt{\omega _ \mathrm{sp} ^ 2 + \qty(\frac{k _ x v}{2}) ^ 2 \pm\omega _ \mathrm{sp} ^ 2\sqrt{e ^ {-2\abs{\bm{k}}L} + \qty(\frac{k _ x v}{\omega _ \mathrm{sp}}) ^ 2}},
\end{align}
where $\omega _ \mathrm{sp} = \omega _ \mathrm{p}/\sqrt{2}$ is the surface plasmon resonance frequency.
As pointed out in previous works~\cite{maslovski2013quantum,brevik2022fluctuational}, the interaction between two surfaces will bring about a growing mode whose eigenfrequency has positive imaginary part $\Im \qty{\omega _ -} \approx \omega _ \mathrm{sp}e ^ {-2\omega _ \mathrm{sp}L/v}/2 := \kappa(L,v)/2$, which increases with the velocity $v$.
Roughly speaking, in our lossy setup,
the overall growth rate in time can be estimated as $[\kappa(L,v) - \gamma]/2$, where the first term arises due to the gain provided by the moving bodies, whereas the second term is due to the material absorption.
When the growth rate exceeds the damping rate, i.e.,
\begin{align}
  e ^ {-2/\bar{v}} - \bar\gamma \gtrsim 0,
  \label{eq:estimate}
\end{align}
the system spontaneously emits light (lasing triggered by the physical motion).
We defined $\bar{v} = v/(\omega _ \mathrm{sp}L)$ and $\bar\gamma = \gamma/\omega _ \mathrm{sp}$ as a normalised velocity and a normalised damping strength, respectively.
From this inequality, it is possible to estimate when the unstable regime is reached.
The critical (normalised) velocity is estimated to be:
\begin{align}
  \bar{v} _ \mathrm{cr} \approx \frac{-2}{\log \bar\gamma}
  \label{eq:estimate_v}
\end{align}
so that the instability condition can be expressed as $\bar{v} > \bar{v} _ \mathrm{cr}$.
To surpass the threshold $\bar{v} _ \mathrm{cr}$, we need either a large $v$ or a small $L$, as both imply $\bar{v} \gg 1$.
For fixed $\gamma$ and $L$, Eq.~\eqref{eq:estimate_v} determines a critical value for the velocity $v _ \mathrm{cr}$ beyond which the system becomes unstable.
Conversely, for fixed $\gamma$ and $v$, the same equation determines a critical value for the distance $L _ \mathrm{cr}$, below which the system also enters the instability regime.

In order to confirm these estimations, next we study the characteristic equation~\eqref{eq:dispersion} in the presence of loss $\gamma > 0$.
The critical values that were discussed earlier are indicative of a crossing of the real-frequency axis. This crossing occurs as one of the roots transitions from the lower half of the frequency plane to the upper half side, so that $\Im \qty{\omega _ -} \approx (\kappa - \gamma)/2 = 0$.
In the following, we analyse the roots of the characteristic equation in the presence of damping to find the critical conditions.

In \figref{fig:poles}, we show the positions of roots for various wavenumbers $k _ x$ in the (i) stable, (ii) critical, and (iii) unstable regimes.
The root in the low-frequency region ($\Re \qty{\omega} \approx 0$) moves towards the imaginary axis as the wavenumber $k _ x$ increases,
 and upon reaching this axis, it ascends along it.
In the case of (ii) critical and (iii) unstable regimes, the root enters the upper-half complex frequency plane, and the corresponding mode will exhibit exponential growth over time, corresponding to instability.

The system is stable if and only if all of the imaginary parts of the roots of the characteristic equation are negative for a real-valued wave vector.
If, for any mode, the corresponding eigenfrequency is located in the upper half plane, the system will no longer reach a steady state, leading to divergence of the frictional force~\cite{brevik2022fluctuational}.
Indeed, for a complex eigenfrequency in the upper half-plane ($\omega = \omega' + i\omega''$ with $\omega'' > 0$), the corresponding mode experiences exponential growth in time and diverges, $e ^ {-i\omega t} \propto e ^ {+\omega'' t} \rightarrow \infty\ (t \rightarrow \infty)$.
In a previous study~\cite{brevik2022fluctuational}, it was shown that such a positive imaginary part of an eigenfrequency (and the resulting divergence) may be brought about by squeezing-type interaction between two modes, which can be regarded as parametric amplification in the quantum system~\cite{scully1997quantum,gardiner2004quantum}.
This instability may be utilised to parametrically amplify surface plasmon fields.

The characteristic equation \eqref{eq:dispersion} has four dimensionless parameters: the wavenumber $\tilde{k} = k _ x/k _ \mathrm{p}$, the damping constant of the metal $\tilde\gamma = \gamma/\omega _ \mathrm{p}$, the sliding velocity $\tilde v = v/c$, and the distance between the two surfaces $\tilde L = k _ \mathrm{p}L$. 
We introduced $k_\mathrm{p} =\omega_\mathrm{p} /c$. The stability condition corresponds to:
\begin{align}
  M = \max _ {\tilde k} \qty{\tilde\omega''(\tilde k, \tilde \gamma, \tilde v, \tilde L) } < 0.
\end{align}
Note that here we take $k_y=0$, because, at the instability threshold, the wave vector is along \textit{x}. 

\begin{figure}[htbp]
  \centering
  \includegraphics[width=\linewidth]{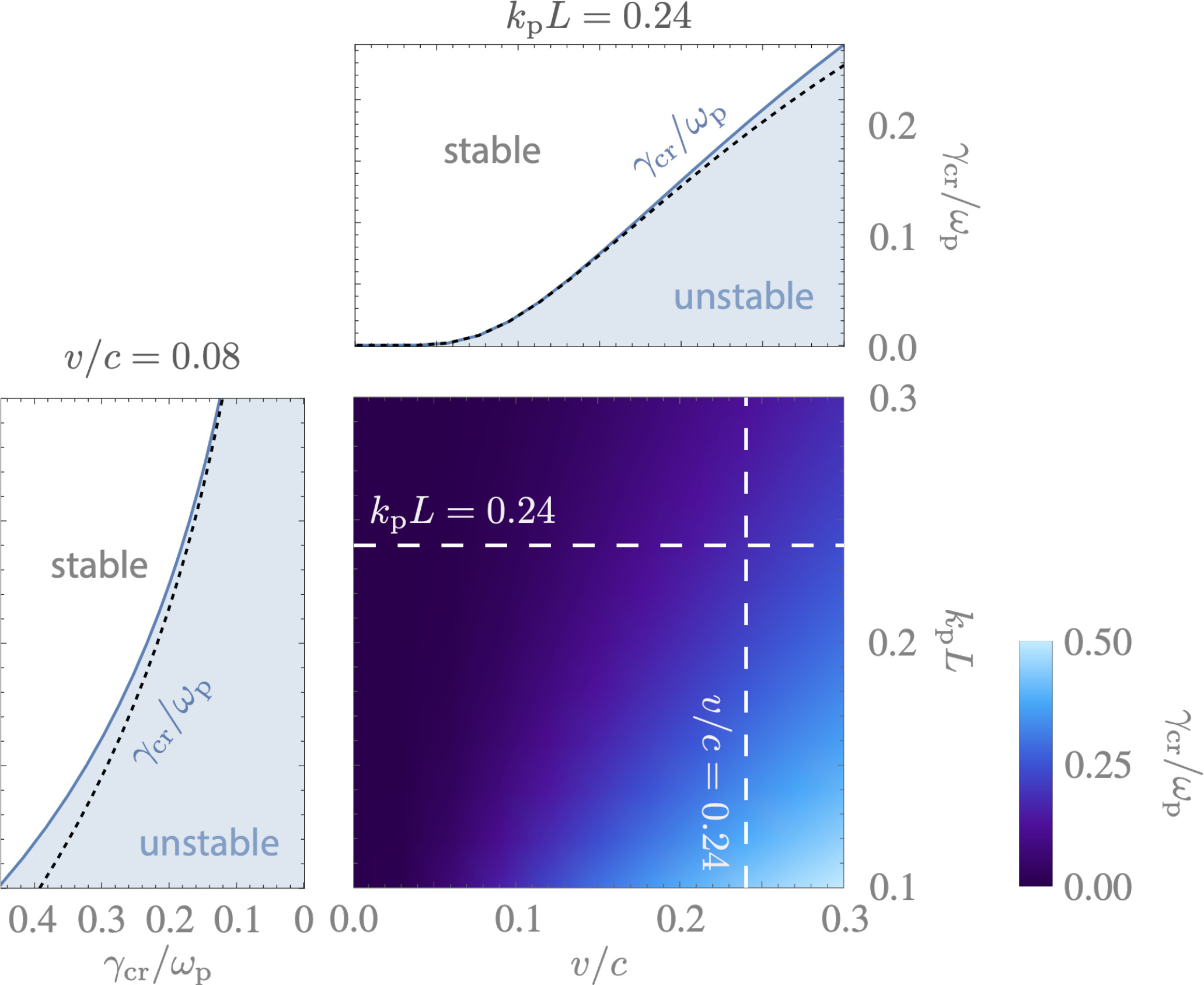}
  \caption{Critical damping strength as a function of the relative velocity $v/c$ and of the distance between two surfaces $k _ \mathrm{p}L$.
    The two plots represent the critical damping evaluated over the two white dashed lines on the colour map.
    The dotted black curves in the two plots depict the analytical estimates [Eq.~\eqref{eq:estimate}], which agree well with the numerical results (solid blue curves) in the low-velocity regime ($v/c \ll 1$) and weak-interaction regime ($k _ \mathrm{p} L \gg 1$) [i.e., the regime where losses dominate].
    If the velocity is large enough, or if the two surfaces are close enough, compared to the damping strength $\gamma/\omega _ \mathrm{p}$, one of the roots of the characteristic equation~\eqref{eq:dispersion} moves to the upper half in the complex frequency plane, leading to instability.
    The threshold is denoted by $\gamma _ \mathrm{cr}$.
    In essence, these two parameters control the interaction between two surfaces.
    If the gain dominates over the damping (exceeds the critical damping strength), plasmonic excitations between the sheared surfaces grow exponentially and trigger instability.
  }
  \label{fig:diagram}
\end{figure}

In \figref{fig:diagram}, we represent the threshold $\tilde \gamma _ \mathrm{cr}$ that leads to $M = 0$.
If the damping rate exceeds the critical value $\tilde \gamma _ \mathrm{cr}$, the system remains stable, and the frictional force can be time-independent (here, we neglect the change in $v$ due to the frictional effect, which is a very slow process).
The system becomes unstable for a damping rate less than $\tilde \gamma _ \mathrm{cr}$ as, in that case, it supports at least one mode with a real-valued wavevector $\bm{k}$ that grows exponentially in time.
It is also worth noting that the analytical estimates [Eq.~\eqref{eq:estimate}] indicated by the black dashed curves in the line plots in  \figref{fig:diagram} agree well with the numerical result if the slabs are well-separated ($k _ \mathrm{p} L \gg 1$) or if the relative velocity is small ($v/c \ll 1$).

%% file: src/noqf.tex
\section{No quantum friction?}
Next, we consider the effect of fluctuation-induced sources.
From the fundamental principles encapsulated with the fluctuation-dissipation theorem, the fluctuating current can be decomposed into positive and negative-frequency parts.
For a lossy system, as Gruner et al.~have shown~\cite{gruner1995correlation,gruner1996green,dung1998three}, the positive-frequency part is written as
\begin{align}
  \bm{j} _ 1 ^ + = \frac{\omega _ 1}{c}\sqrt{\frac{\hbar}{\pi\mu _ 0}\epsilon _ 1''} \bm{f} _ 1,
\end{align}
where $\bm{f} _ 1 := \bm{f} _ {\omega _ 1 \bm{k} _ 1}(z _ 1)$ is a bosonic annihilation operator, satisfying $[{\bm{f} _ 1},{\bm{f} _ 2 ^ \dagger}] = \delta _ {12}$, and $\mu _ 0$ is the vacuum permeability.
The negative-frequency part is given by the Hermitian conjugate, $\bm{j} _ 1 ^ - = (\bm{j} _ 1 ^ +) ^ \dagger$.
The relevant current correlation function will be
\begin{align}
  \expval{\acomm{\bm{j} _ 1 ^ +}{\bm{j} _ 2 ^ -}}
  = \frac{\hbar\omega _ 1 ^ 2\epsilon _ 0}{\pi}\epsilon _ 1''\delta _ {12},
  \label{eq:j1j2_PL}
\end{align}
where the angular brackets $\expval{\ldots}$ represent the vacuum expectation value.
If we naively apply these conventional expressions to our setup (without incorporating the necessary modification described in the next section), the symmetrised field-field correlation will be written purely in terms of the imaginary part of Green's function~\cite{philbin2009no},
\begin{align}
  T _ {12} = 
  \frac{\hbar}{8\pi^3}
  \Im\qty{\bar{G} _ {12} + \bar{G} _ {21} ^ \top},
\end{align}
where we have defined $\bar{G} = (\omega ^ 2/c ^ 2)G$, and the superscript $\top$ denotes the transpose.
The frictional force on the lower surface per unit of area will be given by the $xz$ component,
\begin{align}
  F _ {12} = \lim _ {z _ {1,2} \rightarrow z _ -} \bm{u} _ x \cdot T _ {12} \bm{u} _ z.
  \label{eq:E1E2_PL}
\end{align}
Note that the fields are evaluated in the vacuum region just above the lower surface, $z _ {1,2} > z _ -$.
Below, we show that $F _ {12}$ given by the above formula vanishes. This would lead to a negative result, i.e., imply that the shear motion is not associated with a frictional effect (see also previous discussion~\cite{philbin2009no,pendry2010quantum,leonhardt2010comment,pendry2010reply}).
However, we argue that the above formula is wrong and that the standard fluctuation-dissipation relation is inapplicable to gain media.
The simplest argument to show this is to note that the current-current correlation implicit in Eq.~\eqref{eq:j1j2_PL} gives a nonsensical result in the gain regions.
In fact, when $\epsilon'' <0$, it implies that $\expval{\acomm{\bm{j} _ 1 ^ +}{\bm{j} _ 1 ^ -}} < 0$, which is unphysical and mathematically inconsistent with the definition of the correlation function.
It is also possible to show that the correlations for the fields given by Eq.~\eqref{eq:E1E2_PL} are also mathematically inconsistent with the definition of the correlations when the system has gain (see Appendix \ref{appx:EE} for the details).
The problematic behaviour of the current correlation function stems from the Doppler shift.
Recall that the imaginary parts of our Doppler-shifted permittivities~(\ref{eq:eps_p<0}, \ref{eq:eps_m<0}) and hence the right-hand side of Eq.~\eqref{eq:j1j2_PL} can be negative for sufficiently short wavelengths $\omega < v\abs{\bm{k}}/2$.
In other words, as the noise current is written in terms of the square root of the imaginary part of the permittivity, one needs to be careful in a gain system because the square root originates an imaginary number.

In order to prove that the symmetrised field correlation~\eqref{eq:E1E2_PL} vanishes with a misprescribed fluctuating current, we first introduce a `reciprocal dual' system, where the sliding velocity is reversed ($v \to  - v$).
Note that the dual system can be reached by 180$^\circ$ rotation around the $z$ axis so that we can write Green's function for the dual system,
\begin{align}
  \recG _ {12} = P G _ {\bar{1}\bar{2}}P.
  \label{eq:180-rotation}
\end{align}
where we have defined the 180$^\circ$ rotation matrix, $P := \operatorname{diag}(-1,-1,1)$, and the bar symbol denotes flipping the sign of wavevector [i.e., $G _ {\bar{1}\bar{2}} = G _ {\omega _ 1, -\bm{k} _ 1,\omega _ 2, - \bm{k} _ 2}(z _ 1, z _ 2)$].
Note that we have $-\bm{k}$ instead of $\bm{k}$ on the right-hand side because the rotation about the $z$-axis takes $\bm{x} _ \parallel$ to $-\bm{x} _ \parallel$ and consequently $\bm{k}$ to $-\bm{k}$.
On the other hand, the original and the dual transformed system are related to each other through a reciprocity transformation. Consequently, the Green's functions of the two systems are linked as follows (see Appendix \ref{appx:reciprocity} for the detail):
\begin{align}
  G _ {12} = [\recG _ {\bar{2}\bar{1}}] ^ \top.
  \label{eq:reciprocity}
\end{align}
A reciprocity transformation flips all the time-odd macroscopic quantities that determine the response of a material (in our problem, the velocity).
Equation \eqref{eq:reciprocity} establishes that interchanging the observation and source points and flipping the time-odd parameter leads to qualitatively similar wave effects.
Note that the conventional reciprocity relation is recovered at $v \rightarrow 0$, $G _ {12} = G _ {\bar2\bar1} ^ \top$.
Combining these relations (\ref{eq:180-rotation}, \ref{eq:reciprocity}), we can conclude
\begin{align}
  G _ {12} = [PG _ {21}P] ^ \top.
\end{align}
In particular, it follows that:
\begin{align}
  \bm{u} _ x \cdot G _ {12}\bm{u} _ z = -\bm{u} _ x \cdot G _ {21} ^ \top \bm{u} _ z.
\end{align}
This result demonstrates the right-hand side of Eq.~\eqref{eq:E1E2_PL} vanishes,
$F \propto \bm{u} _ x \cdot \qty(G _ {12} + G _ {21} ^ \top) \bm{u} _ z = 0$. 

%% file: src/qf.tex
\section{Quantum friction}
As we have seen in the previous section, the naive application of the fluctuation-dissipation relation for lossy media leads to unphysical correlation functions and a negative result for `quantum friction'.
As previously discussed, the problem arises due to the negative imaginary part of the permittivity in the low-frequency regime due to the Doppler effect.
The negative imaginary part of the permittivity represents wave amplification (gain) in those frequencies.
A prescription to quantise the electromagnetic field in the presence of gain instead of loss is to take the absolute value of the imaginary part of the permittivity and to swap the roles of annihilation and creation operators~\cite{matloob1997electromagnetic,scheel1998qed,vogel2006quantum,raabe2008qed}.
This procedure is compatible with the input-output formalism~\cite{jeffers1993quantum,jeffers1996canonical,scheel2000entanglement,loudon2003quantum,amooghorban2013quantum}, justified by the path-integral formulation with Glauber's inverted harmonic oscillators~\cite{amooghorban2011casimir}, and has recently been utilised for non-Hermitian photonics with complex geometries in the presence of loss and gain~\cite{allameh2016quantization,akbarzadeh2019spontaneous,franke2021fermi,ren2021quasinormal,franke2022quantized}.
With such a prescription, we can write the source current as follows:
\begin{align}
  \bm{j} _ 1 ^ + = \frac{\omega _ 1}{c}\sqrt{\frac{\hbar}{\pi\mu _ 0}\abs{\epsilon _ 1''}}\qty{\theta[+\epsilon _ 1''] \bm{f} _ 1 + \theta[-\epsilon _ 1''] \bm{f} _ 1 ^ \dagger},
\end{align}
where $\theta$ stands for the Heaviside step function.
The corresponding current-current correlation is
\begin{align}
  \expval{\acomm{\bm{j} _ 1 ^ +}{\bm{j} _ 2 ^ -}} = \frac{\hbar \omega _ 1 ^ 2 \epsilon _ 0}{\pi} \abs{\epsilon _ 1''} \delta _ {12}.
  \label{eq:j1j2_QF}
\end{align}
It evidently determines a positive-definite kernel, in agreement with the properties of the mathematical structure of the correlation function.

Using the above result, it is possible to determine a modified fluctuation-dissipation~\cite{scheel1998qed,raabe2008qed,jeffers1993quantum}. The pertinent symmetrized field-field correlation can be written as:
\begin{align}
  T _ {12}
  &= \frac{\hbar}{8\pi^3}
  \Im\qty{\bar{G} _ {12} + \bar{G} _ {21} ^ \top}
  + \frac{2\hbar}{\pi} \Re \int\limits_{\epsilon _ 3 ''<0}
  \bar{G} _ {13} \abs{\epsilon _ 3 ''} \bar{G} _ {23}^ \dagger \dd{3}
  \label{eq:E1E2-pre}
\end{align}
Note that $\dd{3} = \dd{\omega _ 3}\dd{\bm{k} _ 3}\dd{z _ 3}/4\pi^2$ as we defined before, and the integral is restricted to the domain specified by $\epsilon _ 3 = \epsilon _ {\omega _ 3,\bm{k} _ 3}(z _ 3) < 0$ where the medium behaves as a gain medium.
The first term in Eq.~\eqref{eq:E1E2-pre} is just the familiar fluctuation-dissipation result for passive media. The second term is a correction due to the active nature of the system. As seen in the previous section, the first term does not contribute to the friction force. Taking this into account, one can readily show that the force on the lower surface per unit of area is given by:
\begin{align}
  F _ {12} 
  = 
  \lim _ {z _ {1,2} \rightarrow z _ -} 
  \frac{2\hbar}{\pi} \Re \int\limits_{\epsilon _ 3 ''<0}
  \bm{u} _ x \cdot \bar{G} _ {13} \abs{\epsilon _ 3 ''} \bar{G} _ {23}^ \dagger \bm{u} _ z \dd{3}
  \label{eq:E1E2}
\end{align} 
From this expression, it is clear that the active nature of the medium ($\epsilon _ 3'' < 0$) plays a central role in generating the frictional force.
Note that we have adopted the shorthand notation $F _ {12} := F _ {\omega _ 1, \bm{k} _ 1, \omega _ 2, \bm{k} _ 2}$.
The imaginary part of the permittivity controls the strength of fluctuation contributing to the frictional force, while Green's functions are in charge of propagating the fluctuation from a generic point $z _ 3$ in the medium to the vacuum gap region $z _ 1$.
As we analysed in Eqs.~\eqref{eq:eps_p<0} and \eqref{eq:eps_m<0}, the imaginary part of the permittivity in the upper (lower) medium $z > z _ +$ ($z < z _ -$) becomes negative left-propagating $k _ x < 0$ waves (right-propagating $k _ x > 0$ waves) with sufficiently short wavelengths, $\omega < \abs{k _ x}v/2$.
We can complete the evaluation of the frictional force by performing the integrals in Eq.~\eqref{eq:E1E2} with relevant Green's functions.
To this end, we adopt the quasi-static form of Green's function hereafter and semi-analytically evaluate the integrals (See Appendix \ref{appx:green} for the details about Green's function).
For the contribution from the upper medium ($z _ 1 < z _ + < z _ 3$), we write
\begin{align}
  \bar{G} _ {12} &= 
  \frac{-1}{\epsilon _ {2+}}
  \bm{\mathcal{D}} _ 1 \bm{\mathcal{D}} _ 2 ^ \top g _ {12},
  \label{eq:G12+}
  \\
  g _ {12} &= 
  \frac{\bcancel\delta _ {12}}{2\abs{\bm{k} _ 1}}
  \frac{t _ +\qty(1 + r _ - e ^ {-2\abs{\bm{k} _ 1}z _ 1})}{1 - r _ + r _ -} e ^ {+\abs{\bm{k} _ 1}(z _ 1 - z _ 2)},
\end{align}
while the lower medium contribution ($z _ 3 < z _ - < z _ 1$) is written as
\begin{align}
  \bar{G} _ {12} &= 
  \frac{-1}{\epsilon _ {2-}}
  \bm{\mathcal{D}} _ 1 \bm{\mathcal{D}} _ 2 ^ \top g _ {12},
  \label{eq:G12-}
  \\
  g _ {12} &=
  \frac{\bcancel\delta _ {12}}{2\abs{\bm{k} _ 1}}
  \frac{t _ -(1 + r _ + e ^ {+2\abs{\bm{k} _ 1}z _ 1})}{1 - r _ + r _ -}  e ^ {-\abs{\bm{k} _ 1}(z _ 1 - z _ 2)},
\end{align}
where we have written $\bcancel\delta _ {12} = \delta(\omega _ 1 - \omega _ 2)\delta(\bm{k} _ 1 - \bm{k} _ 2)$ and utilised the reflection coefficient $r _ {+(-)}$ of the upper (lower) surface [defined in Eq.~\eqref{eq:r _ pm}], and the corresponding transmission coefficient $t _ \pm = 2\epsilon _ \pm/(1 + \epsilon _ \pm)$.
Note that the arguments of reflection and transmission coefficients are omitted for simplicity.
Performing the integration over all possible correlations, we can write the total contribution to the friction on the lower surface, 
\begin{align}
  &F = \int F _ {\omega _ 1, \bm{k} _ 1, \omega _ 2, \bm{k} _ 2} \dd{\omega _ 1}\dd{\bm{k} _ 1} \dd{\omega _ 2}\dd{\bm{k} _ 2} \nonumber
  \\
  &= \int \limits _ {\substack{k _ {x} < 0\\ \omega < \tfrac{\abs{k _ {x}}v}{2}}}
  \frac{-\hbar k _ {x}}{4\pi^3}
  r _ -'' \abs{\frac{t _ + e ^ {-\abs{\bm{k}}z _ +}/\epsilon _ {\omega \bm{k}+}}{1-r _ + r _ -} \sqrt{|\epsilon _ {\omega \bm{k}+}''|}} ^ 2 \dd{\omega}\dd{\bm{k}}
  \notag \\
  &+ \int \limits _ {\substack{k _ {x} > 0\\ \omega < \tfrac{\abs{k _ {x}}v}{2}}}
  \frac{+\hbar k _ {x}}{4\pi^3}
  r _ +'' \abs{\frac{t _ - e ^ {+\abs{\bm{k}}z _ -}/\epsilon _ {\omega \bm{k}-}}{1-r _ + r _ -} \sqrt{|\epsilon _ {\omega \bm{k}-}''|}} ^ 2 \dd{\omega}\dd{\bm{k}}.
  \label{eq:F}
\end{align}
Note that the integration range is limited in the short-wavelength regions (i.e., $\omega < \abs{k _ {x}}v/2$).
The detailed derivation of Eq.~\eqref{eq:F} is provided in Appendix~\ref{appx:F}.
In this equation, the term $\hbar k _ {x}$ represents the wave momentum, the imaginary part of the reflection coefficient $r _ \pm''$ gives the density of states of the absorptive surface, and the squared absolute value represents the intensities of waves emitted from the fluctuating currents.

\begin{figure}[htbp]
  \centering
  \includegraphics[width=.95\linewidth]{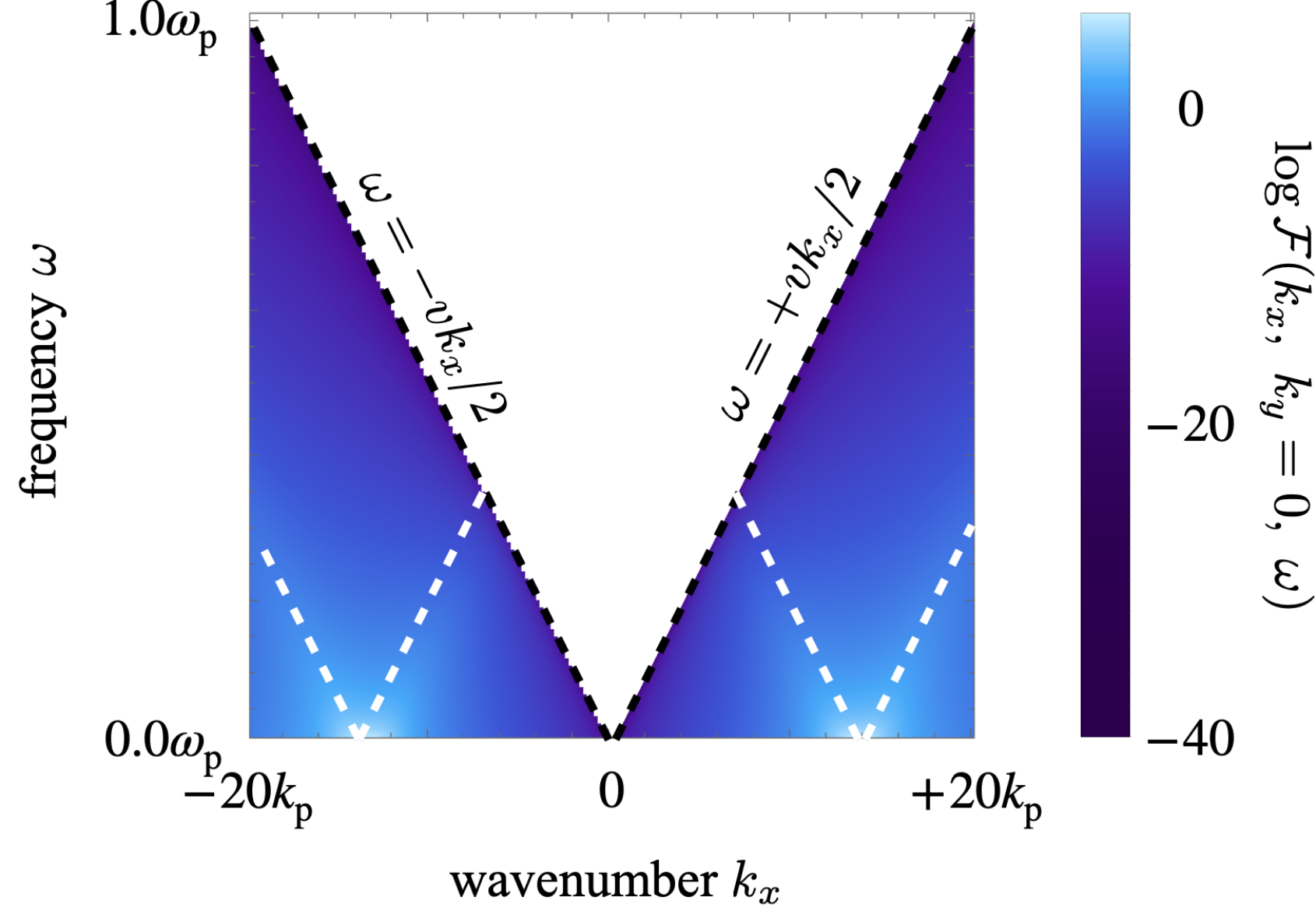}
  \caption{
    Force spectral density.
    Either the upper or the lower region behaves as a gain medium below the black dashed lines $\omega = \abs{k _ x}v/2$.
    In the gain region, the spectral density is peaked in the vicinity of the white dashed lines, which represent Doppler-shifted surface plasmons $\omega = \pm\omega _ \mathrm{sp} \pm k _ x v/2$ ($\omega _ \mathrm{sp} = \omega _ \mathrm{p}/\sqrt{2}$).
    Due to the interaction between the surface plasmons, the peak position slightly deviates from the white dashed lines.
    The following parameters were used to draw this colour map:
    $\gamma = 0.19\omega _ \mathrm{p}$, $v = 0.1c$, and $L = 0.1/k _ \mathrm{p}$ with the plasmon wavenumber $k _ \mathrm{p} = \omega _ \mathrm{p}/c$.
  }
  \label{fig:force-spectrum}
\end{figure}

In \figref{fig:force-spectrum}, we depict the force spectral density [i.e., the integrands in Eq.~\eqref{eq:F}].
In this plot, we set the damping constant slightly above the critical value, $\gamma = 0.19\omega _ \mathrm{p} > \gamma _ \mathrm{cr} \approx 0.18 \omega _ \mathrm{p}$ (for $v = 0.1c$ and $L = 0.1/k _ \mathrm{p}$), so that the system is operated within the stable regime.
In such a case, the spectral force density is free of singularities, allowing for reliable numerical integration.
The magnitude of $\gamma$ adopted above is consistent with value of damping in typical semiconductors~\cite{frigerio2016tunability,panah2017mid,palik1976coupled,gomez2005transmission}.
As seen, the peaks of the force spectrum are located around the Doppler-shifted surface plasmon frequencies (white dashed lines in the figure). 
Note that the interaction between two Doppler-shifted plasmons results in an ``avoided crossing'' so that the peaks are slightly deviated from the dashed lines.
This behaviour stems from the denominators in Eqs.~\eqref{eq:F}, whose roots correspond to elementary excitations (normal modes) in the system.
For the negative wavenumbers $k _ {x} < 0$, only the first contribution in Eq.~\eqref{eq:F} is active, while the second contribution is activated in the positive wavenumber region $k _ {x} > 0$.
The first contribution in Eq.~\eqref{eq:F} represents the process where left-propagating waves are amplified by the upper medium and absorbed by the lower medium, while the second term in Eq.~\eqref{eq:F} corresponds to the process where the lower medium amplifies right-propagating waves which are eventually absorbed by the upper medium.
This is consistent with the discussion in \figref{fig:direction-selective-absorption}.
Importantly, it is shown in Appendix~\ref{appx:consistency} that the formula \eqref{eq:F} is in agreement with previous studies.
It is crucial to underline that, unlike previous works, our approach is not restricted to the quasi-static limit and can be generalized to arbitrary moving systems by adopting a suitable Green's function, which can take relativistic motion into account.
To ascertain the consistency, we can rewrite the transmission coefficients $t _ \pm$ included in Eq.~\eqref{eq:F} in terms of the reflection coefficients $r _ \pm$.
If only the lower slab is moving at the velocity of $v$, only the second term contributes in Eq.~\eqref{eq:F} (see Appendix~\ref{appx:consistency} for the detail),
\begin{align}
    F &\rightarrow \frac{\hbar}{2\pi^3}\int \limits _ {\substack{k _ x > 0\\ \omega < k _ x v}} \frac{k _ x\ r''(\omega)\ r''(\omega - k _ x v)e ^ {-2\abs{\bm{k}}L}}{|1-r(\omega)r(\omega - k _ x v)e ^ {-2\abs{\bm{k}}L}| ^ 2} \dd{\omega}\dd{\bm{k}},
    \label{eq:F_lower-only}
\end{align}
where we defined $r (\omega) := (1 - \epsilon _ \mathrm{D}(\omega))/(1 + \epsilon _ \mathrm{D}(\omega))$.
This representation is consistent with the previous works~\cite{pendry1997shearing,pendry1998can,volokitin1999theory,volokitin2007near}.
In the weak-interaction regime ($k _ \mathrm{p} L \gg 1$), we can approximate the denominator in Eq.~\eqref{eq:F_lower-only} as $|1-r(\omega)r(\omega - k _ x v)e ^ {-2\abs{\bm{k}}L}| ^ 2 \approx 1$.
Moreover, in the lossless limit ($\gamma \rightarrow 0$), the imaginary part of the reflection coefficient becomes the delta function, $r''(\omega) \rightarrow \pi\omega _ \mathrm{sp}\qty{\delta(\omega + \omega _ \mathrm{sp})-\delta(\omega - \omega _ \mathrm{sp})}/2$, as a consequence of the Sokhotski-Plemelj relation.
Therefore, in the weak-interaction regime at the lossless limit, the force acquires the form
\begin{align}
    F \rightarrow -\frac{\hbar \omega _ \mathrm{sp} ^ 3}{4\pi v ^ 2}\int e ^ {-2\sqrt{\qty({2\omega _ \mathrm{sp}}/{v}) ^ 2 + k _ y ^ 2}L}\dd{k _ y}.
\end{align}
This aligns precisely with the result derived in Ref.~\cite{brevik2022fluctuational}.

\begin{figure}[htbp]
  \centering
  \includegraphics[width=\linewidth]{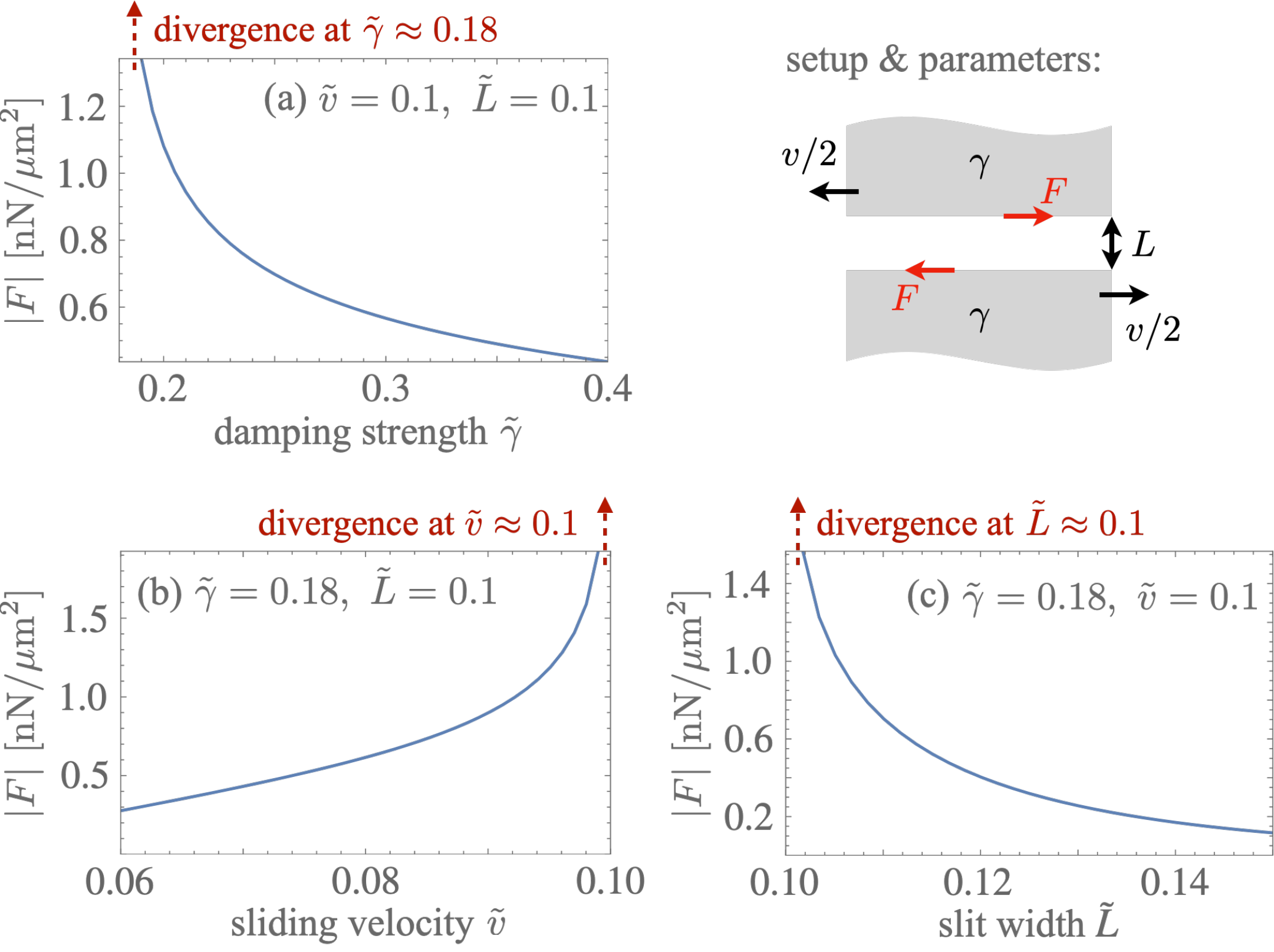}
  \caption{
    Frictional force as a function of (a) the damping strength $\tilde\gamma = \gamma/\omega _ \mathrm{p}$, (b) the sliding velocity $\tilde v = v/c$, and (c) the width of the vacuum slit $\tilde L = k _ \mathrm{p} L$.
    The force diverges if one of the following conditions is fulfilled:
    (a) the damping is sufficiently small,
    (b) the sliding velocity is sufficiently large,
    (c) the vacuum slit width is sufficiently small.
    This is because the system may develop instabilities in those regimes.
    We used the following parameters to draw these figures:
    (a) $\tilde v = 0.1,\ \tilde L = 0.1$; 
    (b) $\tilde\gamma = 0.18,\ \tilde L = 0.1$; 
    (c) $\tilde\gamma = 0.18,\ \tilde v = 0.1$.
  }
  \label{fig:force}
\end{figure}

As the force spectral density is peaked in the short-wavelength region, we can straightforwardly perform the numerical integration of the spectral density in Eq.~\eqref{eq:F} to get the total frictional force per unit of area and study the dependence on the system parameters.
Note that the force spectral density has no singularities in the integration range, provided the classical system is stable (i.e., the eigenfrequencies of the natural modes are in the lower-half frequency plane).
In \figref{fig:force}, we show the total frictional force as a function of (a) the damping strength $\gamma$, (b) the sliding velocity $v$, and (c) the vacuum gap width $L$.
Interestingly, the force increases as the damping becomes weak and diverges at a critical point $\gamma _ \mathrm{cr} \approx 0.18 \omega _ \mathrm{p}$ ($v = 0.1c,\;L = 0.1/k _ \mathrm{p}$).
The divergence of the force occurs when the natural frequencies of the coupled moving slabs cross the real frequency axis and move from the lower-half frequency plane to the upper-half frequency plane.
Beyond the critical point, the system becomes unstable and starts spontaneously emitting light~\cite{silveirinha2014theory,silveirinha2014spontaneous,silveirinha2014optical,lannebere2016negative}, as already discussed in Sec.~III.
At the critical point, there is a phase transition so that the frictional process is no longer stationary (time-independent) but rather grows exponentially in time~\cite{brevik2022fluctuational}.
Thus, we can regard the divergence as a transition from a stable to an unstable regime, where the system will no longer reach steady states.
The force also increases and diverges as the relative velocity $v$ approaches the critical value $v_{\rm{cr}} \approx 0.1c$ ($\gamma = 0.18\omega_\mathrm{p},\;L = 0.1/k_\mathrm{p}$).
This aligns with the rough estimation given in Eq.~\eqref{eq:estimate}.
Previous studies~\cite{guo2014singular,guo2014giant} conjectured that the singular behaviour is logarithmic at the immediate vicinity of the threshold.
Alternatively, the force also diverges as the vacuum gap width $L$ diminishes, reaching the critical value of $L_{\rm{cr}} \approx 0.1/k_\mathrm{p}$ for $\gamma = 0.18\omega_\mathrm{p}$ and $v = 0.1c$.
This is in line with the approximate estimation in Eq.~\eqref{eq:estimate}.

%% file: src/discussion.tex
\section{Discussions}
There is a remarkable parallelism between the instabilities in the quantum friction problem and the Kelvin-Helmholtz instability (KHI).
KHI is one of the most ubiquitous fluid instabilities that occurs if there is a velocity difference across the interface between two fluids.
It was originally studied by Helmholtz~\cite{helmholtz1868discontinuous} and Kelvin~\cite{thomson1871hydrokinetic} and is described by a complex-valued eigenvalue problem within the linearised theory.
The linearised theory consists of the Laplace equation for the fluid pressure $P _ {+(-)}$ above (below) the interface~\cite{chandrasekhar1961hydrodynamic,drazin_2002,charru2011hydrodynamic,schmid2012stability},
\begin{align}
\begin{cases}
    \nabla ^ 2 P _ + = 0 &(z > 0),
  \\
  \nabla ^ 2 P _ - = 0 &(z < 0),
\end{cases}
\label{eq:laplace}
\end{align}
and relevant boundary conditions~\cite{chandrasekhar1961hydrodynamic,drazin_2002,charru2011hydrodynamic,schmid2012stability}.
These are the continuity of the fluid pressure across the interface,
\begin{align}
P _ + = P _ -\quad (z=0),
  \label{eq:P_continuous}
\end{align}
and the kinematic boundary condition,
\begin{align}
  V _ \pm = \qty(\pdv{t} + v _ \pm \pdv{x})\eta\quad(z=0),
  \label{eq:kinematic_bc}
\end{align}
where we defined the upper (lower) fluid velocity $V _ {+(-)}$ in the $z$ direction and the fluid velocity $v _ {+(-)}$ in the $x$ direction, and $\eta$ is a function that determines the fluctuation of the interface profile.
Note that both fluids experience the same displacement $\eta$ with respect to the equilibrium position ($z=0$).
Thus, both $V_+$ and $V_-$ can be expressed in terms of $\eta$ as in Eq.~\eqref{eq:kinematic_bc}, thereby interlinking $V_+$ and $V_-$.
The vertical velocity of the fluid near the interface is given by the velocity of the interfacial variation given by $\mathrm{D}\eta/\mathrm{D}t$.
The convective (material) derivative, $\mathrm{D}/\mathrm{D}t = \pdv*{t} + v _ \pm\hspace{-.2em} \pdv*{x}$, is employed to take into account the streaming motion.
This justifies the form of Eq.~\eqref{eq:kinematic_bc}.

The vertical interface velocity observed by each streaming fluid [left-hand side of Eq.~\eqref{eq:kinematic_bc}] can be linked to the fluid pressure via the Euler equation,
\begin{align}
    \rho _ \pm \qty(\pdv{t} + v _ \pm \pdv{x}) V _ \pm = -\pdv{P _ \pm}{z}.
    \label{eq:euler}
\end{align}
where $\rho _ {+(-)}$ is the fluid density above (below) the interface.

Because of the translational symmetry in the $x$ and $y$ directions, we can work in the spectral domain [e.g., use $\widetilde{P} _ \pm (\omega, k _ x, k _ y)$ instead of $P _ \pm(t,x,y)$] as we did in the case of the quantum friction problem.
The Laplace equation can be written as
\begin{align}
    \begin{cases}
    \displaystyle
    \pdv[2]{z} \widetilde{P} _ + = -k ^ 2 \widetilde{P} _ + &(z > 0),
    \vspace{.5em}
    \\
    \displaystyle
    \pdv[2]{z} \widetilde{P} _ - = -k ^ 2 \widetilde{P} _ - &(z < 0),
\end{cases}
\label{eq:KHI_DE}
\end{align}
where $k ^ 2 = k _ x ^ 2 + k _ y ^ 2$, and the continuity equation for pressure maintains the same form,
\begin{align}
\widetilde{P} _ + = \widetilde{P} _ -\quad (z=0).
  \label{eq:KHI_BC1}
\end{align}
In the spectral domain, the fact that the vertical displacement $\eta$ is identical for both fluids together with the kinematic boundary condition~\eqref{eq:kinematic_bc} leads to $\widetilde{V} _ +/(\omega - k _ x v _ +) = \widetilde{V} _ -/(\omega - k _ x v _ -)$, while the Euler equation \eqref{eq:euler} gives $\rho _ \pm \widetilde{V} _ \pm = (\partial{\widetilde{P} _ \pm}/\partial{z})/(\omega - k _ x v _ \pm)$; hence, we can write
\begin{align}
    \frac{\rho ^ {-1}}{(\omega - k _ x v _ +) ^ 2}\pdv{z}\widetilde{P} _ + = \frac{\rho ^ {-1}}{(\omega - k _ x v _ -) ^ 2}\pdv{z}\widetilde{P} _ -
    \quad (z = 0),
    \label{eq:KHI_BC2}
\end{align}
where we have assumed that the two fluids are identical (i.e., $\rho _ + = \rho _ - = \rho$) for simplicity.
Therefore, the linearised theory of KHI can be reduced to Eqs.~(\ref{eq:KHI_DE}--\ref{eq:KHI_BC2}).
In the following, we show that the instability in the quantum friction setup is governed by analogous equations when the two metallic slabs are sufficiently close to one another ($L\rightarrow 0$).

First, our system is described by the Laplace equation since we are working in the quasi-static regime.
In the upper ($+$) and lower ($-$) media, the electrostatic potentials obey respective Laplace equations.
Working in the spectral domain, we can obtain 
\begin{align}
\begin{cases}
    \displaystyle
    \pdv[2]{z} \phi _ + = -k ^ 2 \phi _ + & (z > 0),
    \vspace{.5em}\\
    \displaystyle
    \pdv[2]{z} \phi _ - = -k ^ 2 \phi _ + & (z < 0),
\end{cases}
  \label{eq:qf_DE}
\end{align}
if the vacuum slit between two metallic slabs is vanishingly narrow ($L\rightarrow 0$).
Second, the electrostatic potential should be continuous across the interface,
\begin{align}
    \phi _ + = \phi _ -
    \quad
    (z = 0).
    \label{eq:qf_BC1}
\end{align}
Third, the continuity of the electric flux density reads
$\epsilon _ + \pdv*{\phi _ +}{z}
= \epsilon _ - \pdv*{\phi _ -}{z}\;(z=0)$.
Inspired by the fact that hydrodynamics theory may appear as a low-energy effective description~\cite{romatschke2010new,dubovsky2012effective,florkowski2018new}, we focus on the low energy ($\omega - k _ x v _ \pm \ll \omega _ \mathrm{p})$ regime.
Under those conditions, the dielectric function can be approximated as $\epsilon _ \pm \approx -\omega _ \mathrm{p} ^ 2/(\omega - k _ x v _ \pm) ^ 2$,
and the flux density continuity gives
\begin{align}
    \frac{\omega _ \mathrm{p} ^ 2}{(\omega - k _ x v _ +) ^ 2}\pdv{z}\phi _ +
    = \frac{\omega _ \mathrm{p} ^ 2}{(\omega - k _ x v _ -) ^ 2}\pdv{z}\phi _ -
    \quad
    (z = 0),
    \label{eq:qf_BC2}
\end{align}
when the damping is significantly weak, $\gamma \rightarrow 0$ (i.e., the setup becomes unstable).
We substitute $v _ - = - v _ + = v/2$ to reproduce our setup.
It is clear that the equations governing KHI (\ref{eq:KHI_DE}--\ref{eq:KHI_BC2}) coincide with the ones for the quantum friction setup (\ref{eq:qf_DE}--\ref{eq:qf_BC2}).
Thus, we have established a precise correspondence between the instability in the quantum friction setup and KHI.
Note that equations~(\ref{eq:qf_DE}--\ref{eq:qf_BC2}) are automatically built in the quasi-static Green's function.

%% file: src/conclusion.tex
\section{Conclusion}
In this study, we have developed a rigorous quantum theory to characterize noncontact quantum friction.
 A pivotal aspect of our analysis hinges on the observation that the optical response of a moving body is typically active. 
Thus, a moving body can, in some conditions, behave as a gain medium.
This effect usually requires interactions mediated by waves with very short wavelengths and is controlled by the Doppler shift.
We have adopted a generalised fluctuation-dissipation relation that takes into consideration the Doppler-induced gain in the electromagnetic field quantization. 
By evaluating the expectation value of the stress tensor on one of the moving surfaces without perturbative approximations, we have derived a frictional force formula, which generalises previous studies.
In particular, we find an excellent agreement between our theory and previous works in the quasi-static limit.

Remarkably, our analysis predicts a phase transition from a stable stationary regime to an unstable regime where the force exhibits exponential growth.
Our theory shows that the quantum friction force diverges at the critical transition point.
We have shown that the instabilities can occur when either of the following conditions are met:
(i) the damping is sufficiently weak;
(ii) the shearing velocity is significantly large;
(iii) the surfaces are sufficiently near to each other.
Our analysis shows that the instabilities are rooted in the interactions of quasi-static plasmons, which may ultimately lead to divergent frictional forces.
The origin of the instabilities has been elucidated by identifying the locus of the Green’s functions poles. Furthermore, we have established a precise
parallelism between the quantum friction problem and the Kelvin-Helmholtz instabilities by carefully examining the corresponding equations of motion.

%% file: src/acknowledgement.tex
\begin{acknowledgments}
  D.O.~is supported by JSPS Overseas Research Fellowship, by the Institution of Engineering and Technology (IET), and by Funda\c{c}\~ao para a Ci\^encia e a Tecnologia and Instituto de Telecomunica\c{c}\~oes under project UIDB/50008/2020.
  J.B.P.~acknowledges support from the Gordon and Betty Moore Foundation.
\end{acknowledgments}

%% file: src/appx_EE.tex
\section{Field correlation function}
\label{appx:EE}
If we proceed with the \textit{unmodified} fluctuation-dissipation relation~\eqref{eq:j1j2_PL},
the source current is given as
\begin{align}
  \bm{j} _ 1 ^ + = \frac{\omega _ 1}{c}\sqrt{\frac{\hbar}{\pi\mu _ 0}\epsilon _ 1''}\bm{f} _ 1,
\end{align}
and we can write the corresponding electric field,
\begin{align}
  \bm{E} _ 1 = i\int G _ {12} \omega _ 2 \mu _ 0 \bm{j} _ 2 ^ + \dd{2} + \mathrm{H.c.}
\end{align}
The correlation function in question is calculated as
\begin{align}
  \expval{\bm{E} _ 1 \bm{E} _ 1} 
  &= \int G _ {12} \omega _ 2 \mu _ 0 \expval{\bm{j} _ 2 ^ +\bm{j} _ 3 ^ -} \omega _ 3 \mu _ 0 G _ {13} ^ \dagger \dd{2}\dd{3},
  \notag \\
  &= \frac{\hbar}{\pi\epsilon _ 0}\int \bar{G} _ {13} \epsilon _ 3 '' \bar{G} _ {13} ^ \dagger \dd{3}.
  \label{eq:E1E1}
\end{align}
We recall the definition $\bar{G} = (\omega ^ 2/c ^ 2) G$.
In the case of a gain system, $\epsilon _ 3'' < 0$, we have
\begin{align}
  \bm{u} ^ * \cdot \bar{G} _ {13} \epsilon _ 3 '' \bar{G} _ {13} ^ \dagger \bm{u} 
  = \epsilon _ 3 '' \abs{\bar{G} _ {13} ^ \dagger \bm{u}} ^ 2 < 0,
\end{align}
where $\bm{u}$ is an arbitrary vector.
This implies that the integrand in Eq.~\eqref{eq:E1E1} may be a negative quantity.
However, the integrand itself must be positive definite because the spectral density of the field-field correlation must have that property.

%% file: src/appx_reciprocity.tex
\section{Reciprocity relation}
\label{appx:reciprocity}
Let us consider time-harmonic classical fields (with a single frequency component) defined either in our original setup or in the corresponding reciprocal dual system, where all time-odd macroscopic parameters are flipped (in our problem, the velocity),
\begin{align}
  &\begin{cases}
    \bm{E} _ {\omega _ 1}(x) = \displaystyle \bm{E} _ {\omega _ 1}(\bm{x} _ \parallel,z) = \int \bm{E} _ 1(z) e ^ {i\bm{k} _ 1\cdot \bm{x} _ \parallel} \dd{\bm{k} _ 1},\vspace{.5em}\\
    \bm{H} _ {\omega _ 1}(x) =\displaystyle \bm{H} _ {\omega _ 1}(\bm{x} _ \parallel,z) = \int \bm{H} _ 1(z) e ^ {i\bm{k} _ 1\cdot \bm{x} _ \parallel} \dd{\bm{k} _ 1},\vspace{.5em}\\
    \bm{j} _ {\omega _ 1}(x) =\displaystyle \bm{j} _ {\omega _ 1}(\bm{x} _ \parallel,z) = \int \bm{j} _ 1(z) e ^ {i\bm{k} _ 1\cdot \bm{x} _ \parallel} \dd{\bm{k} _ 1},
  \end{cases}
  \label{eq:EHj}
  \\
  &\begin{cases}
    \recE _ {\omega _ 1}(x) = \displaystyle \recE _ {\omega _ 1}(\bm{x} _ \parallel,z) = \int \recE _ 1(z) e ^ {i\bm{k} _ 1\cdot \bm{x} _ \parallel} \dd{\bm{k} _ 1},\vspace{.5em}\\
    \recH _ {\omega _ 1}(x) = \displaystyle \recH _ {\omega _ 1}(\bm{x} _ \parallel,z) = \int \recH _ 1(z) e ^ {i\bm{k} _ 1\cdot \bm{x} _ \parallel} \dd{\bm{k} _ 1},\vspace{.5em}\\
    \recj _ {\omega _ 1}(x) = \displaystyle \recj _ {\omega _ 1}(\bm{x} _ \parallel,z) = \int \recj _ 1(z) e ^ {i\bm{k} _ 1\cdot \bm{x} _ \parallel} \dd{\bm{k} _ 1},
  \end{cases}
    \label{eq:EHj_reciprocal}
\end{align}
Note that, in the following, we shall omit the arguments and/or subscripts for conciseness where appropriate. 
The field and current amplitudes on the right-hand sides of Eqs.~(\ref{eq:EHj}, \ref{eq:EHj_reciprocal}) satisfy Maxwell's equations,
\begin{align}
  &\begin{cases}
    \bm{\mathcal{D}} _ 1 \times \bm{E} _ {1} = +i\omega _ 1 \mu _ 0 \bm{H} _ {1},\\
    \bm{\mathcal{D}} _ 1 \times \bm{H} _ {1} = -i\omega _ 1 \epsilon _ 0 \epsilon _ {1} \bm{E} _ {1} + \bm{j} _ {1}.
  \end{cases}
  \label{eq:maxwell}
  \\
  &\begin{cases}
    \bm{\mathcal{D}} _ 1 \times \recE _ {1} = +i\omega _ 1 \mu _ 0 \recH _ {1},\\
    \bm{\mathcal{D}} _ 1 \times \recH _ {1} = -i\omega _ 1 \epsilon _ 0 \rec \epsilon _ {1} \recE _ {1} + \recj _ {1}.
  \end{cases}
  \label{eq:maxwell_reciprocal}
\end{align}
where the permittivity distribution for the reciprocal dual system is $\rec \epsilon _ 1 = \epsilon _ 1 (z _ 1;-v)$, with $\epsilon _ 1(z _ 1; +v)$ being the permittivity of the original system defined in Eq.~\eqref{eq:eps_dist}.
Using Maxwell's equations (\ref{eq:maxwell}, \ref{eq:maxwell_reciprocal}), we can obtain the following relation between the fields and current in the real space [quantities on the left-hand side of Eqs.~(\ref{eq:EHj}, \ref{eq:EHj_reciprocal})],
\begin{align}
  &\nabla \cdot \qty(\bm{E} _ {\omega} \times \recH _ {\omega} - \recE _ {\omega} \times \bm{H} _ {\omega}) 
  = \recE _ {\omega} \cdot \bm{j} _ {\omega} - \bm{E} _ {\omega} \cdot \recj _ {\omega},
  \\
  &0 = \int \qty(\recE _ {\omega} \cdot \bm{j} _ {\omega} - \bm{E} _ {\omega} \cdot \recj _ {\omega}) \dd{\bm{x}},
\end{align}
where we have applied the radiative boundary condition (all fields vanish at $\abs{\bm{x}} \rightarrow \infty$) as in the conventional derivation of the reciprocity relation.
Substituting Eqs.~(\ref{eq:EHj}, \ref{eq:EHj_reciprocal}) and integrating over frequencies, we can get
\begin{align}
  \int \qty(\recE _ {\bar{3}} \cdot \bm{j} _ {3} - \bm{E} _ {3} \cdot \recj _ {\bar3}) \dd{3} = 0,
  \label{eq:reciprocity_int}
\end{align}
where the bar symbol indicates flipping the sign of wavevector [e.g.~$\bm{E} _ {\bar{3}} = \bm{E} _ {\omega _ 3, -\bm{k} _ 3}(z _ 3)$].
Setting the current density,
\begin{align}
  \recj _ 3 =  \bm{u} _ {1} \delta _ {3\bar1},\quad
  \bm{j} _ 3 = \bm{u} _ {2} \delta _ {32},
\end{align}
where $\bm{u} _ {1,2}$ are arbitrary vectors, we can write the electric field in terms of Green's functions for the original system $G$ and the dual one $\recG$,
\begin{align}
  \begin{cases}
    \recE _ {\bar3} 
    = \displaystyle \int \recG _ {\bar3 4} \recj _ 4\dd{4} 
    = \recG _ {\bar3\bar1} \bm{u} _ 1,\vspace{.5em}\\
    \bm{E} _ 3 
    = \displaystyle \int G _ {34} \bm{j} _ 4 \dd{4} 
    = G _ {32} \bm{u} _ 2,
  \end{cases}
\end{align}
and the integral \eqref{eq:reciprocity_int} gives
\begin{align}
  \bm{u} _ 1 \cdot \qty[\recG _ {\bar{2}\bar{1}}] ^ \top \bm{u} _ 2 &= \bm{u} _ 1 \cdot G _ {12} \bm{u} _ 2.
\end{align}
Since $\bm{u} _ {1,2}$ are arbitrary, we can conclude
\begin{align}
  \qty[\recG _ {\bar{2}\bar{1}}] ^ \top &= G _ {12}. \tag{\ref{eq:reciprocity}}
\end{align}

%% file: src/appx_green.tex
\section{Quasistatic Green's function}
\label{appx:green}
In the quasistatic regime, electric and magnetic fields are decoupled, and we can write
\begin{align}
  &\bm{E} _ 1 = -\bm{\mathcal{D}} _ 1 \phi _ 1,
  \label{eq:E_phi}
  \\
  &\bm{\mathcal{D}} _ 1 \times \bm{E} _ 1 = 0,
  \label{eq:faraday_quasistatic}
  \\
  &\bm{\mathcal{D}} _ 1 \cdot \bm{E} _ 1 = \frac{\rho _ 1}{\epsilon _ 1 \epsilon _ 0} = \frac{1}{i\omega _ 1\epsilon _ 1/c ^ 2}\bm{\mathcal{D}} _ 1 \cdot \mu _ 0 \bm{j} _ 1
  \label{eq:gauss}
\end{align}
where we introduced the electrostatic potential $\phi$ and the electric charge density $\rho$ and utilised the continuity equation $-i\omega _ 1 \rho _ 1 + \bm{\mathcal{D}} _ 1 \cdot \bm{j} _ 1 = 0$.
Since Eq.~\eqref{eq:E_phi} automatically satisfies Eq.~\eqref{eq:faraday_quasistatic}, we can focus on the third equation~\eqref{eq:gauss}.
We introduce a Green's function for the Poisson equation, defined as
\begin{align}
  -\bm{\mathcal{D}} _ 1 ^ 2 g _ {12} = \delta _ {12}.
\end{align}
Then, the electric field can be written as:
\begin{align}
  &\bm{E} _ 1
  = -\bm{\mathcal{D}} _ 1 \int g _ {12} \frac{cZ _ 0}{i\omega _ 2 \epsilon _ {2}}\bm{\mathcal{D}} _ 2 \cdot \bm{j} _ 2 \dd{2}, \nonumber
  \\
  &= \int \qty{\frac{-1}{\omega _ 2 ^ 2/c ^ 2}\bm{\mathcal{D}} _ 1 \bm{\mathcal{D}} _ 2 ^ \top \frac{g _ {12}}{\epsilon _ 2}} i\omega _ 2\mu _ 0 \bm{j} _ 2\dd{2},
\end{align}
where we performed the integration by parts and used the vacuum impedance $Z _ 0 := \sqrt{\mu _ 0/\epsilon _ 0}$.
Comparing this equation with Eq.~\eqref{eq:E=Gj}, we find that the Green's function can be expressed as:
\begin{align}
  G _ {12} = \frac{-1}{\omega _ 2 ^ 2/c ^ 2}\bm{\mathcal{D}} _ 1 \bm{\mathcal{D}} _ 2 ^ \top \frac{g _ {12}}{\epsilon _ 2}.
\end{align}
Solving the Poisson equation in each region and imposing the field continuity conditions at the surfaces, we can write Green's function $g(z _ 1, z _ 2)$ with the help of reflection and transmission coefficients.
For the field in the vacuum gap coming from the upper medium ($z _ - < z _ 1 < z _ + < z _ 2$), we have
\begin{align}
  g _ {12} = 
  \frac{\bcancel\delta _ {12}}{2\abs{\bm{k} _ 1}}
  \frac{t _ +\qty(1 + r _ - e ^ {-2\abs{\bm{k} _ 1}z _ 1})}{1 - r _ + r _ -} e ^ {+\abs{\bm{k} _ 1}(z _ 1 - z _ 2)},
\end{align}
for the one from the lower medium ($z _ 2 < z _ - < z _ 1 < z _ +$),
\begin{align}
  g _ {12} =
  \frac{\bcancel\delta _ {12}}{2\abs{\bm{k} _ 1}}
  \frac{t _ -(1 + r _ + e ^ {+2\abs{\bm{k} _ 1}z _ 1})}{1 - r _ + r _ -}  e ^ {-\abs{\bm{k} _ 1}(z _ 1 - z _ 2)},
\end{align}
where we defined $\bcancel\delta _ {12} = \delta(\omega _ 1 - \omega _ 2)\delta(\bm{k} _ 1 - \bm{k} _ 2)$ and the reflection coefficient $r _ {+(-)}$ of the upper (lower) surface and the transmission coefficient $t _ {+(-)}$ of the upper (lower) surface.
Note that the propagation factor is included in the coefficients.
Namely, at the quasi-static limit, the reflection coefficients acquire the form, 
\begin{align}
  r _ \pm = \frac{1-\epsilon _ \pm}{1 + \epsilon _ \pm}e ^ {\mp2 \abs{\bm{k}}z _ \pm}.
\end{align}
On the other hand, the transmission coefficient reads
\begin{align}
  t _ \pm = \frac{2\epsilon _ \pm}{1 + \epsilon _ \pm}.
  \label{eq:t=1+r}
\end{align}

%% file: src/appx_F.tex
\section{Derivation of Eq.~\eqref{eq:F}}
\label{appx:F}

From Eqs.~\eqref{eq:eps_p<0} and \eqref{eq:eps_m<0}, the upper (lower) slab [i.e., $z > z _ +\;(z < z _ -)$] behaves as a gain medium, $\epsilon'' _ {+(-)} < 0$, for $k _ x < 0$ ($k _ x > 0$) in the short-wavelength regime $\omega < \abs{k _ x} v/2$.
Thus, we can split the integral $\Re \int \ldots \dd{3} = \Re \int \ldots \dd{\omega _ 3} \dd{\bm{k} _ 3} \dd{z _ 3}/4\pi^2$ in Eq.~\eqref{eq:E1E2} into two parts so that we have the upper and the lower slab contributions $F _ {12} = F _ {12+} + F _ {12-}$,
\begin{align}
    F _ {12+}&= \lim _ {z _ {1,2} \rightarrow z _ -} \frac{2\hbar}{\pi}\Re \int \limits _ {z _ +} ^ {+\infty} \int\limits_{\substack{k _ {3x} < 0 \\ \omega < \tfrac{\abs{k _ {3x}}v}{2}}}
  \bm{u} _ x \cdot \bar{G} _ {13} \abs{\epsilon _ 3 ''} \bar{G} _ {23}^ \dagger \bm{u} _ z \dd{3},
  \label{eq:F12+}
  \\
  F _ {12-} &= \lim _ {z _ {1,2} \rightarrow z _ -} \frac{2\hbar}{\pi}\Re \int \limits _ {-\infty} ^ {z _ -} \int\limits_{\substack{k _ {3x} > 0 \\ \omega < \tfrac{\abs{k _ {3x}}v}{2}}}
  \bm{u} _ x \cdot \bar{G} _ {13} \abs{\epsilon _ 3 ''} \bar{G} _ {23}^ \dagger \bm{u} _ z \dd{3}.
  \label{eq:F12-}
\end{align}
The first and second terms in Eq.~\eqref{eq:F} are derived from Eqs.~\eqref{eq:F12+} and \eqref{eq:F12-}, respectively.
They represent contributions to the friction arising from the correlation between two modes labelled ``1'' and ``2''.

Upon integrating all conceivable correlations following the substitution of the quasi-static form of Green's function~\eqref{eq:G12+}, we can write the overall contribution to the friction from the upper slab,
\begin{align}
    &F _ + = \int F _ {12+} \dd{\omega _ 1}\dd{\bm{k} _ 1} \dd{\omega _ 2}\dd{\bm{k} _ 2}
    \notag \\
    &=\Re \iint \limits _ {z _ +} ^ {+\infty}
     \frac{\hbar}{\pi} \abs{\frac{t _ +/\epsilon _ {3+}}{1 - r _ + r _ -}\sqrt{\abs{\epsilon _ {3+} ''}}} ^ 2
    \notag\\
    &\hspace{4em} \times ik _ {3x}(1 + r _ - e ^ {-2\abs{\bm{k} _ 3} z _ -})e ^ {+\abs{\bm{k} _ 3}(z _ - - z _ 3)}
    \notag\\
    &\hspace{4em} \times \qty{+\abs{\bm{k} _ 3}(1 - r _ - ^ * e ^ {-2\abs{\bm{k} _ 3} z _ -})e ^ {+\abs{\bm{k} _ 3}(z _ - - z _ 3)}} \dd{3},
    \notag
\end{align}
\begin{align}
    &=-\iint \limits _ {z _ +} ^ {+\infty} 
     \frac{\hbar k _ {3x}}{\pi} \abs{\frac{t _ +/\epsilon _ {3+}}{1 - r _ + r _ -}\sqrt{\abs{\epsilon _ {3+} ''}}} ^ 2
     2r'' _ - \abs{\bm{k} _ 3}e ^ {-2\abs{\bm{k} _ 3}z _ 3} \dd{3},
     \notag \\
    &=-\int 
     \frac{\hbar k _ {3x}}{4\pi^3} r'' _ - \abs{\frac{t _ + e ^ {-\abs{\bm{k} _ 3}z _ +} /\epsilon _ {3+}}{1 - r _ + r _ -}\sqrt{\abs{\epsilon _ {3+} ''}}} ^ 2
     \dd{\omega _ 3}\dd{\bm{k} _ 3}.
     \label{eq:F+}
\end{align}
This is the first term (upper-slab contribution) in Eq.~\eqref{eq:F}.
Note that the integration limits for the wavevector and the frequency have been suppressed for brevity.

Analogously, the total contribution to the friction from the lower slab can be evaluated by using the quasi-static Green's function~\eqref{eq:G12-}:
\begin{align}
    &F _ - = \int F _ {12-} \dd{\omega _ 1}\dd{\bm{k} _ 1} \dd{\omega _ 2}\dd{\bm{k} _ 2}
    \notag \\
    &=\Re \iint \limits _ {-\infty} ^ {z _ -}
     \frac{\hbar}{\pi} \abs{\frac{t _ -/\epsilon _ {3-}}{1 - r _ + r _ -}\sqrt{\abs{\epsilon _ {3-} ''}}} ^ 2
    \notag\\
    &\hspace{4em} \times ik _ {3x}(1 + r _ + e ^ {+2\abs{\bm{k} _ 3} z _ -})e ^ {-\abs{\bm{k} _ 3}(z _ - - z _ 3)}
    \notag\\
    &\hspace{4em} \times \qty{-\abs{\bm{k} _ 3}(1 - r _ + ^ * e ^ {+2\abs{\bm{k} _ 3} z _ -})e ^ {-\abs{\bm{k} _ 3}(z _ - - z _ 3)}} \dd{3},
    \notag \\
    &=\iint \limits _ {-\infty} ^ {z _ -} 
     \frac{\hbar k _ {3x}}{\pi} \abs{\frac{t _ -/\epsilon _ {3-}}{1 - r _ + r _ -}\sqrt{\abs{\epsilon _ {3-} ''}}} ^ 2
     2r'' _ + \abs{\bm{k} _ 3}e ^ {+2\abs{\bm{k} _ 3}z _ 3} \dd{3},
     \notag \\
    &=\int 
     \frac{\hbar k _ {3x}}{4\pi^3} r'' _ + \abs{\frac{t _ - e ^ {+\abs{\bm{k} _ 3}z _ -} /\epsilon _ {3-}}{1 - r _ + r _ -}\sqrt{\abs{\epsilon _ {3-} ''}}} ^ 2
     \dd{\omega _ 3}\dd{\bm{k} _ 3}.
     \label{eq:F-}
\end{align}
This is the second term (lower-slab contribution) in Eq.~\eqref{eq:F}.

%% file: src/appx_consistency.tex
\section{Consistency with the previous result~\cite{pendry1997shearing,pendry1998can,volokitin1999theory,volokitin2007near}}
\label{appx:consistency}
We can write the imaginary parts of the reflection coefficients \eqref{eq:r _ pm} as
\begin{align}
  r _ \pm'' = \frac{-2\epsilon _ \pm''}{(1 + \epsilon _ \pm)(1 + \epsilon _ \pm ^ *)}e ^ {\mp 2\abs{\bm{k}}z _ \pm}.
\end{align}
Comparing this representation with the expression of the transmission coefficient \eqref{eq:t=1+r}, we can rewrite the squared amplitudes in Eq.~\eqref{eq:F} in terms of the imaginary part of the reflection coefficient,
\begin{align}
  \abs{\frac{t _ \pm e ^ {\mp \abs{\bm{k}}z _ \pm}}{\epsilon _ {\pm}}\sqrt{|\epsilon _ {\pm}''|}} ^ 2 
  = \frac{-4\epsilon _ {\pm}''e ^ {\mp 2\abs{\bm{k}}z _ \pm}}{(1 + \epsilon _ {\pm})(1 + \epsilon _ {\pm} ^ *)} 
  = 2r _ {\pm}'',
\end{align}
where we have included the exponential factor in the reflection coefficients, consistently with Eq.~\eqref{eq:r _ pm}.
Note that we have used $\epsilon _ {\pm}'' < 0$, which is the range of integration in question, to write $\abs{\epsilon _ {\pm}''} = -\epsilon _ {\pm}''$.
Therefore, the overall friction force takes the form
\begin{align}
    F &=\frac{-\hbar}{2\pi^3}\int \limits _ {\substack{k _ {x} < 0\\ \omega < \tfrac{\abs{k _ {x}}v}{2}}} \frac{k _ x\ r _ -''\ r _ +''}{|1-r _ + r _ -| ^ 2} \dd{\omega}\dd{\bm{k}}
  \notag\\
    &+\frac{+\hbar}{2\pi^3}\int \limits _ {\substack{k _ {x} > 0\\ \omega < \tfrac{\abs{k _ {x}}v}{2}}} \frac{k _ x\ r _ +''\ r _ -''}{|1-r _ + r _ -| ^ 2} \dd{\omega}\dd{\bm{k}}.
    \label{eq:F simplified}
\end{align}
If only the lower slab is moving at a speed of $v$, only the second term in Eq.~\eqref{eq:F simplified} contributes, where we can substitute
\begin{align}
r _ + &\rightarrow r (\omega) e ^ {-2\abs{\bm{k}}z _ +},\\
r _ - &\rightarrow r (\omega - k _ x v)e ^ {2\abs{\bm{k}}z _ -},
\end{align}
with $r (\omega) := (1 - \epsilon _ \mathrm{D}(\omega))/(1 + \epsilon _ \mathrm{D}(\omega))$, and we can obtain
\begin{align}
    F &\rightarrow \frac{\hbar}{2\pi^3}\int \limits _ {\substack{k _ x > 0\\ \omega < k _ x v}} \frac{k _ x\ r''(\omega)\ r''(\omega - k _ x v)e ^ {-2\abs{k}L}}{|1-r(\omega - k _ x v)r(\omega)e ^ {-2\abs{k}L}| ^ 2} \dd{\omega}\dd{\bm{k}}.
    \tag{\ref{eq:F_lower-only}}
\end{align}
This is identical to the previous result~\cite{pendry1997shearing,pendry1998can,volokitin1999theory,volokitin2007near}.